\documentclass[11pt,palatino]{article}
\usepackage[margin=0.75in]{geometry}

\usepackage[numbers,sort&compress,square]{natbib}

\usepackage{amsthm, amsmath, mathtools, mathrsfs}

\usepackage{amsfonts,amsmath,amssymb,bm,url}

\usepackage{graphicx,epsfig,subfigure}

\newcommand{\beginappendix}{%
        \setcounter{table}{0}
        \renewcommand{\thetable}{S\arabic{table}}%
        \setcounter{figure}{0}
        \renewcommand{\thefigure}{S\arabic{figure}}%
     }

\usepackage[utf8]{inputenc} 
\usepackage[T1]{fontenc}    
\usepackage{hyperref}       
\usepackage{url}            
\usepackage{graphicx}
\usepackage{booktabs}       
\usepackage{amsfonts}       
\usepackage{nicefrac}       
\usepackage{microtype}      
\usepackage{xcolor}         
\usepackage{comment}
\usepackage{amsmath}
\usepackage{amsthm}
\usepackage[utf8]{inputenc}
\usepackage[english]{babel}
\usepackage{authblk}
\usepackage{algorithm}
\usepackage{algpseudocode}

\title{Distance by de-correlation: Computing distance with heterogeneous grid cells}

\author[*, 1, 2, 8]{Pritipriya Dasbehera}
\author[*, 3, 4, 5, 8]{Akshunna S. Dogra}
\author[6, 7]{William T. Redman}

\affil[1]{School of Physical Sciences, National Institute of Science Education and Research, India}
\affil[2]{Homi Bhaba National Institute, India}
\affil[3]{NSF AI Institute for Artificial Intelligence and Fundamental Interactions (IAIFI), USA}
\affil[4]{Center for Doctoral Training in Mathematics of Random Systems: Analysis, Modelling and Algorithms (MRS CDT), Imperial College London \& Oxford University, UK}
\affil[5]{Center of Mathematical Sciences and Applications (CMSA), Harvard University, USA}
\affil[6]{Department of Electrical and Computer Engineering, Johns Hopkins University, USA}
\affil[7]{Data Science and AI Institute, Johns Hopkins University, USA}
\affil[8]{MathePhysics}
\affil[*]{Equal Contribution}
\affil[ ]{Correspondence: \texttt{adogra@nyu.edu, wredman4@jh.edu}}

\begin{document}

\maketitle

\begin{abstract}
Encoding the distance between locations in space is essential for accurate navigation. Grid cells, a functional class of neurons in medial entorhinal cortex, are believed to support this computation. Inspired by recent work finding populations of grid cells to have small, but robust heterogeneity in their grid properties, we hypothesize that distance coding can be achieved by a simple de-correlation of population activity. We develop a mathematical theory for describing this de-correlation in one-dimension, showing that its predictions are consistent with simulations of noisy grid cells. Our simulations highlight a non-intuitive prediction of such a distance by de-correlation framework. Namely, some further distances are better encoded than some nearer distances. We find preliminary evidence of this ``sweet spot'' in previously published rodent behavioral experiments and demonstrate that a decoder which estimates distance from the de-correlation of populations of simulated noisy grid cells leads to a similar pattern of errors. Finally, we extend our theory to two-dimensions and, by simulating noisy grid cells in two-dimensions, find that there exists a trade-off between the range of distances that can be encoded by de-correlation of population activity and the distinguishability between different distances, which is controlled by the amount of variability in grid properties. We show that the previously measured average amount of grid property variability strikes a balance, enabling the encoding of distances up to several meters. Our work provides new insight on how grid cells can underlie the encoding of distance and why grid cells may have small amounts of heterogeneity in their grid properties. 
\end{abstract}

\section*{Author Summary}
Grid cells, a functional class of neurons in medial entorhinal cortex that exhibit a periodic distribution of their firing fields in space, are believed to play an integral role in supporting spatial navigation. Recent work has found that grid cells within the same anatomical region (``grid module'') have small, but robust variability in their grid patterns. Such heterogeneity has generally been assumed not to exist in theoretical work exploring the computational role of grid cells. Here, we show that this variability, which on the face of it appears to present difficulties for reconciling grid cells role in spatial navigation, can enable a simple and efficient code for measuring distance between points in space. We develop a mathematical theory to explain how grid property variability shapes this code and simulate noisy grid cell populations to explore trade-offs between distinguishability between distances and the range of distances that can be coded. Preliminary analysis of neurophysiological recordings and behavioral experiments supports aspects of the theory. Our work demonstrates that grid property variability is not a limitation for grid cell involvement in spatial navigation, but may provide considerable capability.

\section{Introduction}
Grid cells, a functional class of neurons found in medial entorhinal cortex (MEC) \cite{hafting2005microstructure, stensola2012entorhinal}, are canonically viewed as providing a metric for space, encoding the distance between points \cite{fiete2008grid, schoyen2025hexagons,xu2025on} and enabling path integration \cite{mcnaughton2006path, burak2009accurate, sorscher2023unified} and spatial navigation \cite{bush2015using, edvardsen2020navigating}, more generally. This presumed computational role is supported by experimental work showing that distortions in grid firing patterns are correlated with errors in distance estimation \cite{duncan2025grid}. Furthermore, lesioning or inhibiting MEC leads to a significant increase in distance estimation error \cite{jacob2017medial, tennant2018stellate}. 

Theoretical work and analysis of neurophysiological data have shown that, up to the distance of half a grid spacing (up to $\approx$ $1$ m), distance can be decoded from grid cells through their population activity (referred to as the ``population vector'') \cite{schoyen2025hexagons, xu2025on, chaudhuri2025not}. In particular, the greater the distance between two points, the larger the distance between corresponding population vectors. This demonstrates that, on a limited spatial scale, grid cells provide a conformal isometry \cite{xu2022conformal}. To encode distances beyond half a grid spacing, which is necessary for spatial navigation of real-world environments, a more complex neural code has been hypothesized \cite{fiete2008grid, sreenivasan2011grid}. In particular, the quasi-periodic relationship between the firing fields of grid cells from different grid modules, which differ in their intrinsic grid spacing and orientation \cite{stensola2012entorhinal}, generates a code with high capacity \cite{fiete2008grid, sreenivasan2011grid}. This code has been shown to be robust to noise \cite{sreenivasan2011grid} and the observed progression of increasing grid spacing across modules can been explained by normative arguments \cite{wei2015principle, stemmler2015connecting}. How downstream brain regions can extract this information, especially when they receive inputs from a minority of the grid modules \cite{van2009anatomy}, is unclear. 

Recent work has shown that individual grid modules contain grid cells with not a single grid spacing and orientation, as has often been assumed in theoretical analysis, but a small, yet robust variability in their grid properties \cite{redman2025robust}. Importantly, if this heterogeneity is a general feature of the grid code, then previous theoretical works \cite{fiete2008grid, burak2009accurate, sreenivasan2011grid, mathis2012optimal, wei2015principle, stemmler2015connecting} -- which assumed grid cells to be identical, up to translation -- do not \textit{a priori} hold. From a computational perspective, this appears to be a steep price to pay, given that this theory attributes the grid code with several powerful properties. A natural question to ask then is, what computational properties does a grid code with variability in grid spacing and orientation possess?

As a first attempt to answer this, Redman et al. (2025) showed that the heterogeneity in grid spacing and orientation enables \textit{single} grid modules to encode local spatial information \cite{redman2025robust}, a computation which previously had been thought to require the integration of activity across multiple grid modules \cite{mathis2012optimal, wei2015principle, stemmler2015connecting}. However, there are several reasons why this was unsatisfying. First, to achieve a high level of accuracy in spatial decoding, simulations showed that hundreds of grid cells were needed. Second, greater accuracy in decoding of local space was achieved with variability that was greater than what was found in analysis of the neural recordings. And third, it has been shown that non-grid cells in the MEC can contain robust firing patterns \cite{ismakov2017grid}, which could be leveraged to further improve decoding accuracy of spatial position \cite{stefanini2020distributed}. Taken together, these results question the optimality of a grid code with heterogeneity for encoding local space. Of course, the biological brain may not ``care'' about such optimality, and other considerations may lead to variability in grid spacing and orientation. However, inspired by the many hypothesized ways in which properties of neural codes enable powerful computations, we ask, are there any other benefits grid property heterogeneity could confer to the grid code? 

\begin{figure}
    \centering
    \includegraphics[width=0.85\linewidth]{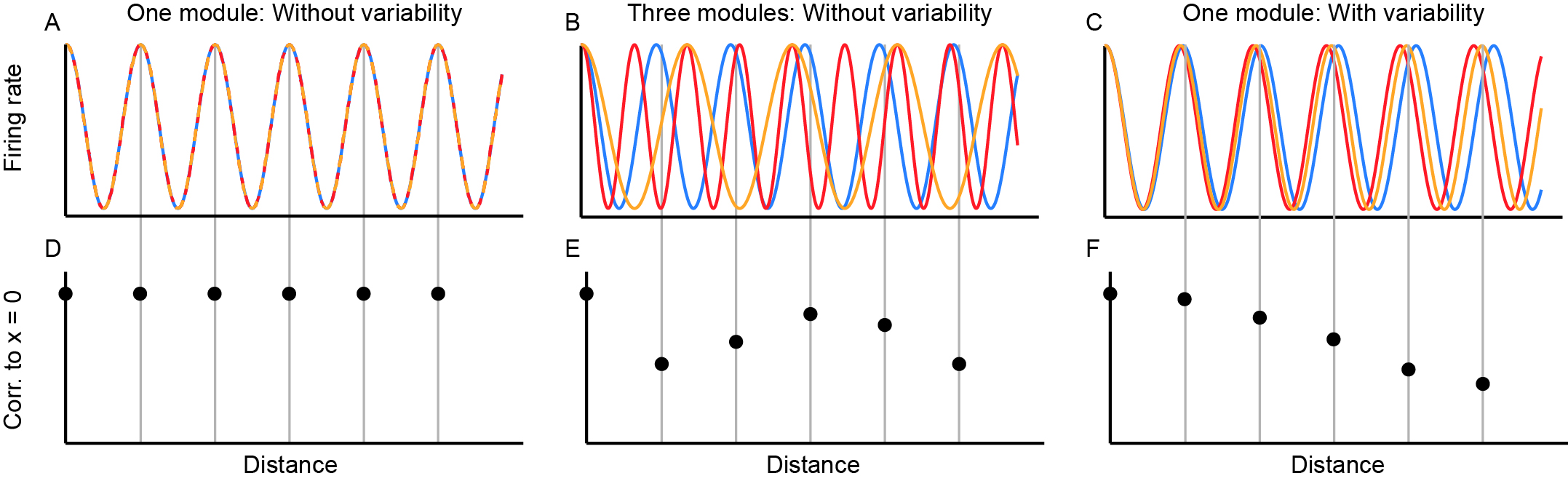}
    \caption{\small \textbf{Schematic of how variability in grid spacing can lead to monotonic relationship between distance and population vector correlation.} (A)--(C) Example idealized 1D firing fields for grid cells coming from: one module, with no variability in grid spacing (A); three modules, with no variability in grid spacing (B); and one module, with variability in grid spacing (C). (D)--(F) Example correlation between population vectors corresponding to distances of integer multiples of the grid spacing with the population vector corresponding to the starting location, for the same populations of grid cells as (A)--(C). }
    \label{fig:schematic}
\end{figure}

Motivated by the putative role of grid cells for distance coding, we hypothesize that grid property variability could enable a simple and efficient way to achieve this (Fig. \ref{fig:schematic}). If there is no variability in grid properties (Fig. \ref{fig:schematic}A), then the correlation of the population vector, at any distance that is an integer multiple of the grid spacing, with the population vector at some reference location ($x = 0$) is constant (Fig. \ref{fig:schematic}D). This is due to the degeneracy in grid firing fields, as every grid cell in the module is the same, up to a translation. If activity from grid cells across multiple modules, each without any variability in grid properties, is considered (Fig. \ref{fig:schematic}B), a symmetry breaking occurs and the population vector, at integer multiples of the grid spacing, has a unique correlation with the population vector at $x = 0$ (Fig. \ref{fig:schematic}E). However, because the grid spacing of different grid modules is often incommensurate, there is not a monotonic relationship between distance and correlation, although nonlinear decoding methods could nevertheless enable accurate estimation of distance \cite{fiete2008grid, sreenivasan2011grid}. With this coding scheme, a smaller correlation \textit{does not} imply a greater distance. In contrast, if grid cell activity from a single grid module with non-zero variability in grid properties is considered, not only is there again a symmetry breaking (Fig. \ref{fig:schematic}C), but population vectors at integer multiples of the grid spacing have correlations with the population vector at $x = 0$ that monotonically decay with distance (Fig. \ref{fig:schematic}F). Thus, in this case, a smaller correlation \textit{does} imply a greater distance. 

Here, we examine the hypothesis of grid cells enabling a ``distance by de-correlation'' coding scheme in detail. First, we analyze previously recorded MEC activity in rodents running along a one-dimensional (1D) environment \cite{campbell2021distance}. We find preliminary evidence of the grid population vector decaying approximately monotonically with distance (Sec. \ref{subsection: Grid cell correlation decays}). While the data has limitations, preventing a full conclusion to be made, it motivates a greater interrogation of how a distance by de-correlation code could be generated. To address this, we develop a theory for how variability in grid properties affects the population vector correlation of grid cells, as a function of distance, in 1D environments (Sec. \ref{subsection: Grid variability 1D}). We provide numerical support for the validity of this theory through the simulation of noisy grid cells. Next, we highlight a prediction made by the simulations on the fidelity with which populations of heterogeneous grid cells encode distance. Namely, the error in distance encoding should increase, then decrease, and then increase again (Sec. \ref{subsection: Sweet spot}). We find preliminary experimental support of this ``sweet spot'' in prior work that trained mice to travel specific distances \cite{tennant2018stellate}, although future work is needed to assess the extent to which this observation holds. We furthermore show that decoding distance from simulated noisy grid cells can produce similar results to what was found experimentally. We then demonstrate that a thresholded rate-based model enables downstream neurons to encode distance at integer multiples of the grid spacing by summing over heterogeneous grid cell inputs and applying a threshold (Sec. \ref{subsection: Thresholded ratebased model}). This provides intuition for how such a distance by de-correlation code could be implemented in the brain. Finally, we extend our theory to two-dimensions (2D) and perform numerical simulations in 2D spaces (Fig. \ref{subsection: Grid variability 2D}). We find that, as in the 1D case, there exists a trade-off between the fidelity with which distances can be encoded through de-correlation and the total distance over which the de-correlation occurs. This trade-off is controlled by the amount of variability in grid properties. Simulating many combinations allows us to identify that the average heterogeneity in grid spacing and orientation identified previously \cite{redman2025robust} strikes a balance between the two ends of the trade-off. 

Collectively, our results demonstrate that small variability in grid properties enables a simple code of distance, further illuminating the computational properties enabled by a heterogeneous grid module. 

\section{Results}
\label{section: Results}

\subsection{Grid cell correlation decays with distance in experimentally recorded MEC data}
\label{subsection: Grid cell correlation decays}

\begin{figure}
    \centering
    \includegraphics[width=0.85\linewidth]{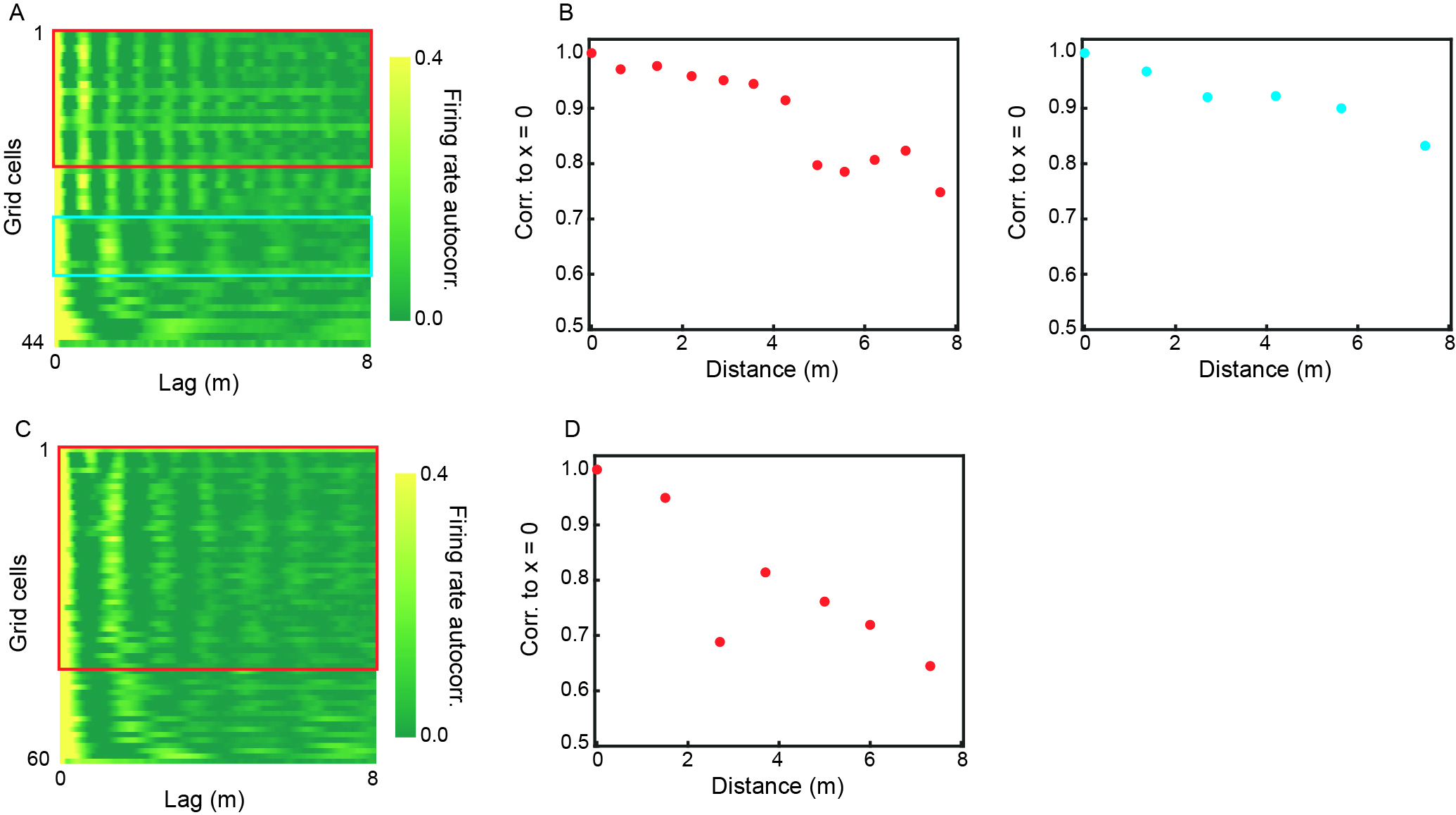}
    \caption{\small \textbf{Preliminary evidence of population vector correlation decaying as a function of distance.} (A) Autocorrelation of firing rate, as a function of lag in distance for all units classified as distance-tuned from one of Campbell et al.'s MEC recordings (recording ID: \texttt{npI5\_0417\_dark\_1}) \cite{campbell2021distance}. (B) Population vector correlation, as a function of distance, when considering only the distance-tuned cells highlighted by the red and blue boxes in (A). These groupings are done to primarily include neurons that have similar spacings between the peaks in their autocorrelations. (C) Same as (A) but for another recording from Campbell et al. (recording ID: \texttt{npI5\_0414\_dark\_1}). (D) Population vector correlation, as a function of distance, when considering only the distance-tuned cells highlighted by the red box in (C). }
    \label{fig:1D_distance_coding_Campbell_data}
\end{figure}

Before we begin tackling the role heterogeneity in grid spacing and orientation may play in enabling a distance by de-correlation code, we start by asking, is there evidence of the kind of monotonic relationship between distance and population vector correlation (schematically illustrated in Fig. \ref{fig:schematic}F) in experimentally recorded grid cells? In order to answer such a question, it is necessary to consider neural activity from MEC while an animal is traversing large distances. One of the few datasets that satisfies this is the work by Campbell et al. (2021), which used Neuropixel probes \cite{jun2017fully} to record from MEC as mice ran over a hundred meters along a 1D virtual track \cite{campbell2021distance}. Campbell et al. (2021) found neurons in MEC that exhibit periodicity in their autocorrelations when running in the dark (Fig. \ref{fig:1D_distance_coding_Campbell_data}A, C), analogous to the periodic structure found in 1D grid cells \cite{gu2018map}. Cells that were found to have sufficient structure in their autocorrelations were considered ``distance-tuned'' \cite{campbell2021distance}.

Inspired by Chaudhuri et al. (2025) \cite{chaudhuri2025not}, we analyzed subsets of the distance-tuned neurons and computed the population vector correlation as a function of distance (Fig. \ref{fig:1D_distance_coding_Campbell_data})(see Sec .\ref{subsection: analysis of 1d grid cell data}, for more details). Importantly, because the autocorrelation decays with distance (Fig. \ref{fig:1D_distance_coding_Campbell_data}A, C), we consider the population vector correlation normalized by the norm of the population vector at each of the two positions (i.e., the cosine similarity) to ensure that we do not see a trivial decay in correlation. When plotting the peaks of population vector correlations (a proxy for integer multiples of the grid spacing), we find it can decay with distance (Fig. \ref{fig:1D_distance_coding_Campbell_data}B, D). This relationship is noisy and not truly monotonic, but there is a general trend towards decreasing population vector correlation with increasing distance. While more detailed analysis and experimental recordings are needed to more completely characterize the extent to which this decay in population vector correlation with distance exists, we believe this result motivates the theoretical and computational results we now turn our attention to.

\subsection{Grid property variability and distance coding in 1D}
\label{subsection: Grid variability 1D}

We develop a theory for how heterogeneity in grid properties affects the population level autocorrelation function. To make the analysis tractable and consistent with the analyzed neural data, we consider a 1D environment. 

We model grid cell activity as a sum of Gaussians, each centered along integer multiples of the grid spacing, 
\begin{equation}
\label{eq: grid cell equation}
    G_{\lambda, \hspace{0.25mm} \omega}(x) = \sum_{n \in \mathbb{Z}} \exp\left(\frac{-(x - n\lambda)^2}{2\omega^2}\right),
\end{equation}
where $\lambda$ is the grid spacing and $\omega$ is the width of the Gaussian. In addition to grid spacing, grid cells are described by their grid orientation, $\theta$, and their grid phase, $\phi$ (e.g., the translation along the $x-$axis of the grid fields). We further simplify by assuming $\theta = 0^\circ$ and $\phi = 0$ m. This is consistent with some experimental work showing a locking of grid fields to the starting location in 1D environments \cite{campbell2021distance}  (Fig. \ref{fig:1D_distance_coding_Campbell_data}A, C), however it makes our analytical theory a conservative lower-bound on the amount of de-correlation that occurs.

Eq. \ref{eq: grid cell equation} is equivalent to a convolution of a Dirac comb with a Gaussian, 
\begin{equation}
    \label{eq: grid cell equation dirac comb}
    G_{\lambda, \hspace{0.25mm} \omega}(x) = \operatorname{comb}_\lambda (x)* g_\omega (x),
\end{equation}
where $g_\omega(x) = \exp(-x^2 / 2\omega^2)$ and $*$ is the convolution operator. In this form, the Fourier transform becomes,
\begin{equation}
    \label{eq: fourier grid cell activity}
    \widehat{G}_{\lambda, \hspace{0.25mm} \omega}(\xi) = \widehat{\operatorname{comb}}_\lambda(\xi) \cdot \widehat{g}_\omega(\xi).
\end{equation}
Eq. \ref{eq: fourier grid cell activity} can be further realized as 
\begin{equation}
\label{eq: expanded fourier grid cell activity}
   \widehat{G}_{\lambda, \hspace{0.25mm} \omega}(\xi) \propto  \sum_{k \in\mathbb Z} \delta \left(\xi - \frac{2\pi k}{\lambda}\right) \cdot e^{-\frac{1}{2} \omega^2\xi^2},
\end{equation}
where $\delta(\cdot)$ is the Dirac delta operator. The power spectrum of $\widehat{G}_{\lambda, \hspace{0.25mm} \omega}(\xi)$ is therefore given by 
\begin{equation}
    \label{eq: power spectrum single grid cell}
    \left|\widehat{G}_{\lambda, \hspace{0.25mm} \omega}(\xi)\right|^2 \propto  \sum_{k \in\mathbb Z} \delta \left(\xi - \frac{2\pi k}{\lambda}\right) \cdot e^{-\omega^2\xi^2}.
\end{equation}

By the Wiener--Khinchin theorem, the autocorrelation, $R_{\lambda, \hspace{0.25mm} \omega}(r)$, is the inverse Fourier transform of the power spectrum \cite{Ricker2003-ye}. In real space, one convenient representation is 
\begin{equation}
    \label{eq: autocorrelation}
    R_{\lambda, \hspace{0.25mm} \omega}(r) \propto \operatorname{comb}_\lambda(r) * e^{-r^2/(4\omega^2)} = \sum_{n \in\mathbb Z} e^{-\frac{(r - n\lambda)^2}{4\omega^2}}. 
\end{equation}
That is, the autocorrelation is a sum of Gaussians, each with width $\sqrt{2} \omega$. This is in agreement with the fact that autocorrelation of a Gaussian with width $\omega$ is another Gaussian of width $\sqrt{2}\omega$.

We now consider a population of $N$ grid cells, each with their own grid spacing, $\lambda_i$. We denote the collection of spacings by $\Lambda = \{\lambda_i\}_{i = 1}^N$. We will assume that they all have a fixed width, $\omega$. We can consider the average population activity as 
\begin{equation}
    \label{eq: population grid cell activity}
    G_{\Lambda, \hspace{0.25mm} \omega}(x) = \frac{1}{N}\sum_{i = 1}^N G_{\lambda_i, \hspace{0.25mm} \omega}.
\end{equation}
If each $\lambda_i$ is sampled from a distribution, $\mathcal{D}$, then we can consider the ensemble averaged (or expected) value, $\mathbb{E}_\mathcal{D} [G_{\Lambda, \hspace{0.25mm} \omega}(x)]$. As the Fourier transform is a linear operation, we can take the expectation on $\widehat{G}_{\Lambda, \hspace{0.25mm} \omega}(\xi)$.

From the above formulation, we can see that a lack of variability in grid spacing (i.e., $\lambda_i = \lambda$, for $i = 1,...,N$), which is assumed in the ``traditional'' perspective of grid cells, leads the population average activity (Eq. \ref{eq: population grid cell activity}) to be the same as a single grid cell (as all grid cells are the same). Therefore, the autocorrelation (Eq. \ref{eq: autocorrelation}) exhibits periodicity that does not decay (Fig. \ref{fig:schematic}A, D) and cannot be used for encoding information about distance. However, if we consider a population of grid cells with variability in grid spacing (like was found in neurophysiological recordings \cite{redman2025robust}), and we assume that $\lambda_i \sim \mathcal{N}(\lambda, \sigma^2_\lambda)$, then the harmonic frequencies of the Fourier transform, $\xi_k(\lambda) = 2\pi k / \lambda$, get shifted. For small variability ($\sigma_\lambda << \lambda$), a first-order Taylor series expansion gives 
\begin{equation}
\label{eq: taylor series}
\xi_k(\lambda_i) \approx \xi_k(\lambda) + \frac{d\xi_k}{d\lambda} \Big|_\lambda (\lambda_i - \lambda),
\end{equation}
where $d\xi_k / d\lambda =-2\pi k / \lambda^2$.

Hence, to leading order,
\begin{equation}
\xi_k(\lambda_i) \sim \mathcal{N} \left(\mu_k,\sigma_k^2\right),
\end{equation}
where $\mu_k = 2\pi k / \lambda$ and $\sigma_k = 2\pi |k| \sigma_\lambda / \lambda^2$. Thus, each spectral line at $\xi_k(\lambda)$ becomes broadened into a Gaussian of width $\sigma_k$. Averaging over the distribution of $\lambda_i$ therefore yields 
\begin{equation}
    \label{eq: expected fourier transform}
    \mathbb{E}_{\lambda_i} \left[\widehat{G}_{\lambda, \hspace{0.25mm} \omega} (\xi) \right] \propto \sum_{k \in \mathbb Z} \exp\left(\frac{-(\xi - \mu_k)^2}{2\sigma_k^2}\right) \cdot e^{-\frac{1}{2} \omega^2\xi^2}.
\end{equation}

Although the summation is over $k \in \mathbb{Z}$, not all $k$ significantly contribute, due to the exponential decay \( e^{-\frac{1}{2} \omega^2\xi^2} \). The \(k\) that most strongly contributes must be such that \(\omega\xi \approx \omega\mu{_{k}} \) is small. This means \(\frac{\omega k}{\lambda} \) is also small and \( \sigma_k << 2 \pi / \lambda \). The expected power distribution, in the limit of $\sigma_k << 2 \pi / \lambda$ (i.e., $\sigma_\lambda << \lambda$), becomes 
\begin{equation}
    \label{eq: expected power distribution}
    \left|\mathbb{E}_{\lambda_i} \left[\widehat{G}_{\lambda, \hspace{0.25mm} \omega} (\xi) \right] \right|^2 \propto \sum_{k \in \mathbb Z} \exp\left(\frac{-(\xi - \mu_k)^2}{2\sigma_k^2}\right)^2 \cdot e^{-\omega^2\xi^2} \approx \sum_{k \in \mathbb Z} \exp\left(\frac{-(\xi - \mu_k)^2}{\sigma_k^2}\right) \cdot e^{-\omega^2\xi^2}.
\end{equation}

From Eq. \ref{eq: expected power distribution}, we see that our theory makes several predictions on the expected power distribution, and hence, on the expected autocorrelation of the activity of population of grid cells with variable grid spacing. In particular, we note that since $\sigma_k = 2\pi |k| \sigma_\lambda / \lambda^2$, Eq. \ref{eq: expected power distribution} predicts that $\left|\mathbb{E}_{\lambda_i} \left[\widehat{G}_{\lambda, \hspace{0.25mm} \omega} (\xi) \right] \right|^2$ will\footnote{A rigorous derivation that yields these predictions is presented in Sec. \ref{subsection: Exact derivation of population vector correlation}.}:
\begin{itemize}
    \item (\textbf{P1}) Smoothly decay with increased distance, as $\sigma_k \propto k$;
    \item (\textbf{P2}) Exhibit an increased rate of decay with increased variability in spacing, as $\sigma_k \propto \sigma_\lambda$;
    \item (\textbf{P3}) Exhibit a decreased rate of decay with increased grid spacing, since $\sigma_k \propto \frac{1}{\lambda^2}$.
\end{itemize}

We test these predictions by simulating populations of noisy grid cells and examining the population vector correlation along increasing integer multiples of the grid spacing. In particular, we sample idealized grid cell ratemaps, constructed using standard approaches \cite{solstad2006grid}, and use them as the means of a Poisson process (see Sec. \ref{subsection: synthetic grid cell simulations}, for more details). 

\begin{figure}
    \centering
    \includegraphics[width=0.875\linewidth]{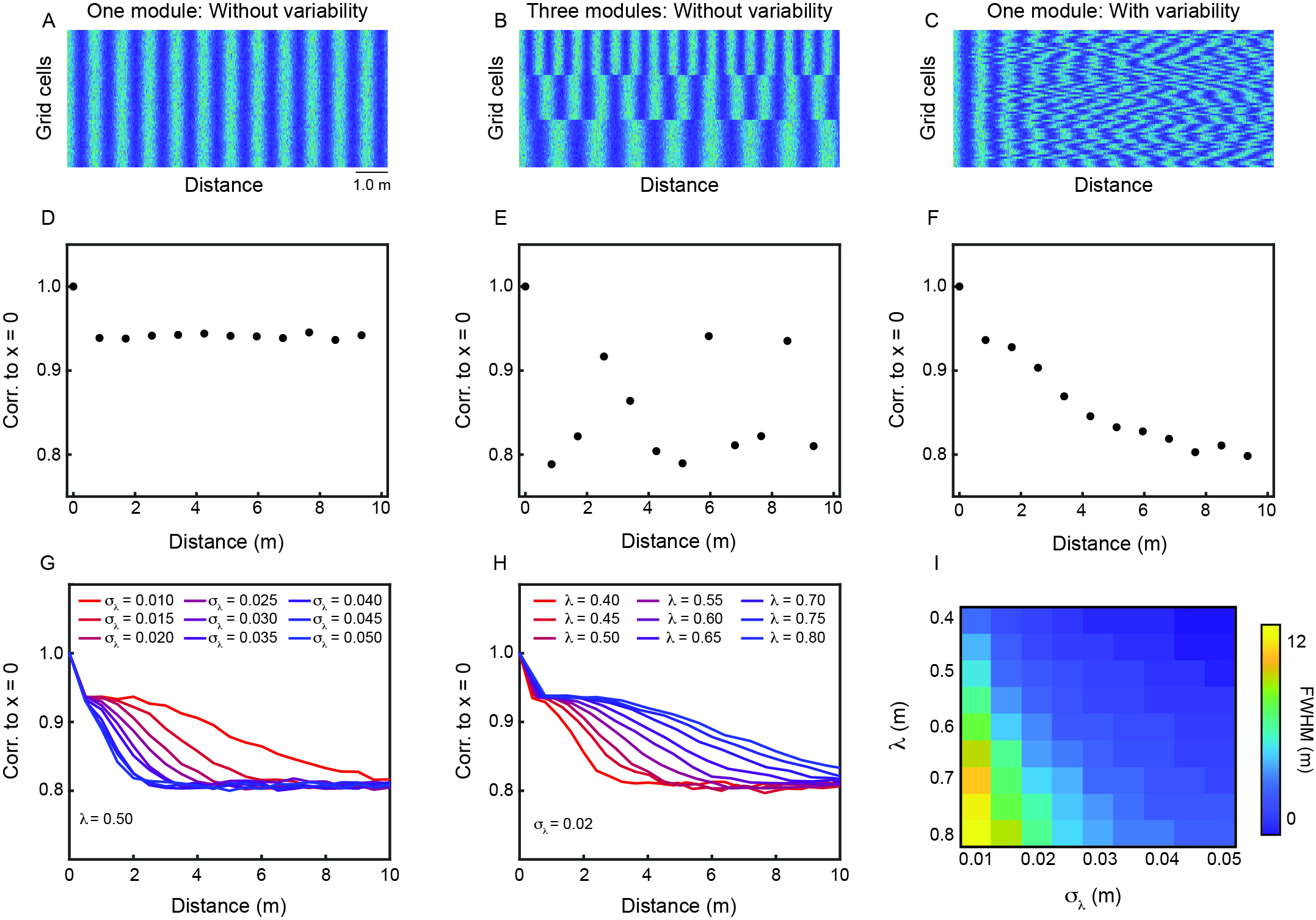}
    \caption{\small \textbf{Variability in grid spacing leads to monotonic relationship between population vector correlation and distance in 1D.} (A)--(C) Noisy grid cell ratemaps in 1D for populations comprising of: a single module, with no variability in grid spacing (A) ($\lambda = 0.85$ m and $\sigma_\lambda = 0$ m); three modules, with no variability in grid spacing (B) ($\lambda_1 = 0.60$ m, $\lambda_2 = 0.85$ m, $\lambda_3 = 1.20$ m, and $\sigma_{\lambda_m} = 0$ m); and one module, with variability in grid spacing (C) ($\lambda = 0.85$ m and $\sigma_\lambda = 0.05$ m). For all populations, the grid orientation is set to $0^\circ$ and the phase of all grid cells is set to $0$ m. (D)--(F) Correlation of population vectors, corresponding to distances of integer multiples of the grid spacing, with the population vector corresponding to the starting location, for the same populations of grid cells as (A)--(C). (G) Population vector correlation, as a function of distance, for different grid spacing variabilities, $\sigma_\lambda$. $\lambda$ is fixed at $0.5$ m. (H) Population vector correlation, as a function of distance, for different grid spacings, $\lambda$. $\sigma_\lambda$ is fixed at $0.02$ m. (I) The full width at half maximum (FWHM) of the population vector correlation curves, for different combinations of $\lambda$ and $\sigma_\lambda$. For all plots, $N = 64$ and $25$ distinct populations were simulated. }
    \label{fig:1D_distance_coding}
\end{figure}

We simulate three types of grid cell populations: grid cells coming from a single module with no variability in grid spacing ($\sigma_\lambda = 0$ m) (Fig. \ref{fig:1D_distance_coding}A); grid cells coming from three modules, each having their own spacing -- $\lambda_m$, which are related by $\lambda_{m + 1} = \sqrt{2} \lambda_m$, which is consistent with previous experimental findings \cite{stensola2012entorhinal} -- and having no variability in grid spacing ($\sigma_{\lambda_m} = 0$ m) (Fig. \ref{fig:1D_distance_coding}B); grid cells coming from a single module with variability in grid spacing ($\sigma_\lambda = 0.05$ m) (Fig. \ref{fig:1D_distance_coding}C). Consistent with \textbf{P1}, we find that the population vector correlation smoothly decays as a function of distance for the population of grid cells with non-zero $\sigma_\lambda$ (Fig. \ref{fig:1D_distance_coding}F). This is not observed for the other two populations (Fig. \ref{fig:1D_distance_coding}D, E), demonstrating that this is a property unique to grid cell populations with heterogeneous grid spacing. 

We then fix $\lambda$ and vary $\sigma_\lambda$. Consistent with \textbf{P2}, we find that increasing $\sigma_\lambda$ leads to an accelerated decay in the population vector correlation (Fig. \ref{fig:1D_distance_coding}G). For instance, when $\sigma_\lambda = 0.01$ m, the population vector correlation does not decay to its baseline until almost $10$ m, exhibiting near monotonic decrease in correlation from $2$ m to $10$ m. In contrast, when $\sigma_\lambda = 0.04$ m, the population vector correlation decays to baseline by $2$ m. When we fix $\sigma_\lambda$ and vary $\lambda$, we find that increasing $\lambda$ leads to slower decay of population vector correlation (Fig. \ref{fig:1D_distance_coding}H), in-line with \textbf{P3}. Simulating a wide range of pairs of $\lambda$ and $\sigma_\lambda$, we find that \textbf{P2} and \textbf{P3} are robustly supported (Fig. \ref{fig:1D_distance_coding}I), demonstrating that our simplified theory can provide insight that is consistent with noisy populations of grid cells. 

\subsection{Existence of sweet spot in distance coding in 1D}
\label{subsection: Sweet spot}

The population vector correlation as a function of distance curves computed through our numerical simulations demonstrate that, after the initial decrease in correlation when moving from the starting location ($x = 0$) to the first grid spacing ($x = \lambda$), the correlation has an approximately sigmoidal dependence on distance (Fig. \ref{fig:1D_distance_coding}G, H). As a consequence of this, the difference in correlation between neighboring points is not uniform. More precisely, if the population vector correlation function has a sigmoidal form 
\begin{equation}
\label{eq: correlation function sigmoid}
    C(x) \propto \frac{1}{1 + e^{-\beta (x - \nu)}},
\end{equation}
then the derivative of the correlation function is
\begin{equation}
\label{eq: derivative correlation function sigmoid}
    \frac{dC(x)}{dx} \propto \frac{-\beta e^{-\beta (x - \nu)}}{\left[ 1 + e^{-\beta (x - \nu)}\right]^2},
\end{equation}
where $\beta$ controls the slope of the sigmoid, $\nu$ determines the midpoint of the sigmoid, and $x = n \lambda$, $n \in \mathbb{N}$. From Eq. \ref{eq: derivative correlation function sigmoid}, we see that the maximum value of $dC(x) / dx$ is $\beta / 4$, which occurs at $x = \nu$. This predicts that there exists a ``sweet spot'', where the difference in correlation between neighboring integer multiples of the grid spacing is maximal and thus, the distinguishability of distance based on correlation is increased. A more rigorous derivation of the sweet spot from the developed 1D theory can be found in Sec. \ref{subsection: Deriving the sweet spot}.

\begin{figure}
    \centering
    \includegraphics[width=0.85\linewidth]{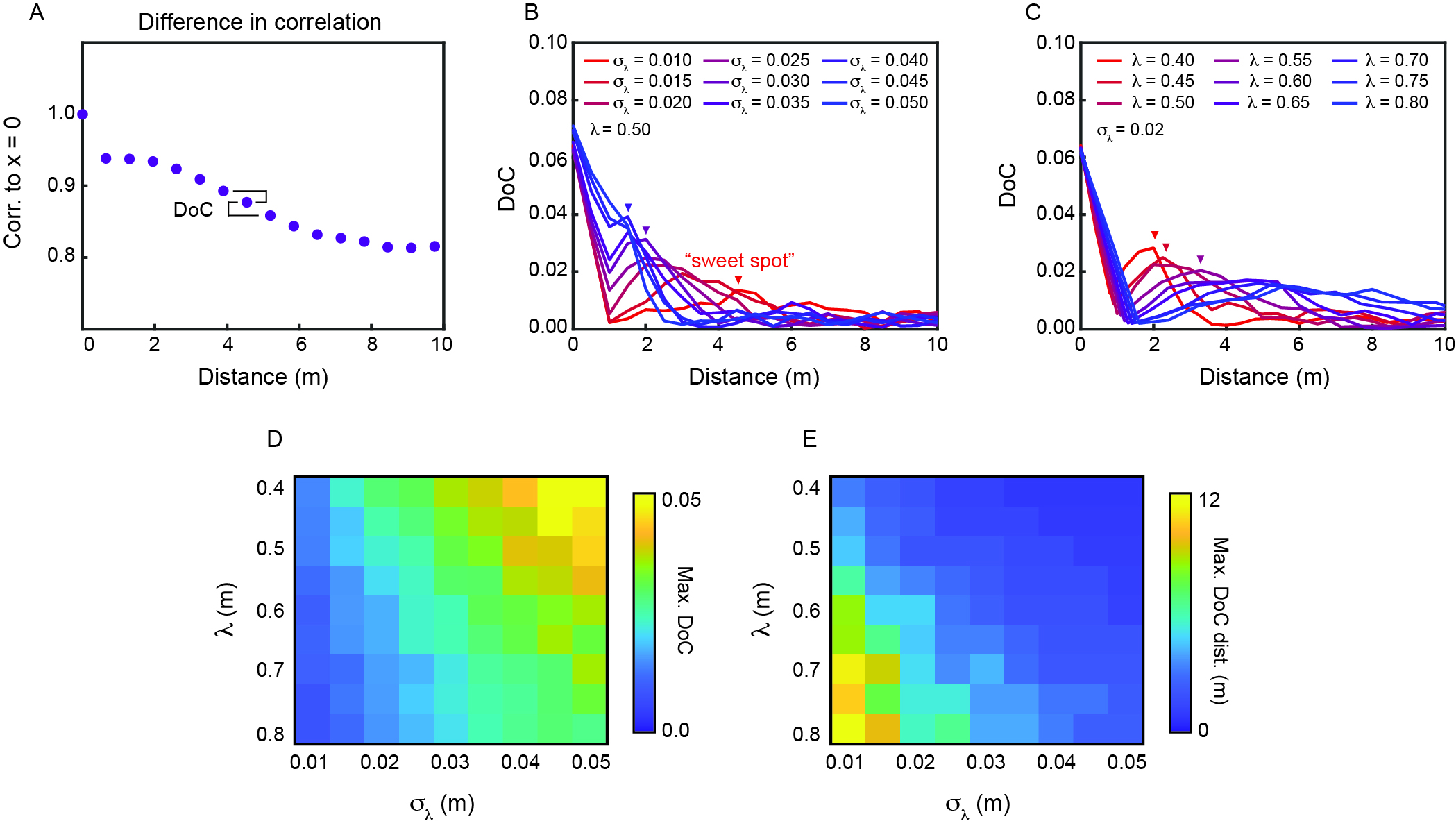}
    \caption{\small \textbf{Grid spacing and grid spacing variability control trade-off between range over which de-correlation occurs and distinguishability between neighboring distances.} (A) Schematic illustration of the difference of correlation (DoC) for one example distance. (B) DoC, as a function of distance, for different grid spacing variabilities, $\sigma_\lambda$. $\lambda$ is fixed at $0.5$ m. (C) DoC, as a function of distance, for different grid spacings, $\lambda$. $\sigma_\lambda$ is fixed at $0.02$ m. (D) Maximum DoC value for different combinations of $\lambda$ and $\sigma_\lambda$. (E) Distance at which  the maximum DoC value occurs, for different combinations of $\lambda$ and $\sigma_\lambda$. For all plots, $N = 128$ and $25$ unique populations were simulated. }
    \label{fig:1D_sweet_spot}
\end{figure}

To test this prediction, we first measure the finite difference in correlation (DoC) of the simulated $C(x)$ curves (Fig. \ref{fig:1D_sweet_spot}A). That is, 
\begin{equation}
\label{eq: DoC}
    \text{DoC}(n \lambda) = \frac{1}{2} \Big( \big|C[n \lambda] - C[(n - 1) \lambda]\big| + \big|C[n \lambda] - C[(n + 1) \lambda]\big| \Big). 
\end{equation}
As predicted by Eq. \ref{eq: derivative correlation function sigmoid}, we find that the DoC has a peak (Fig. \ref{fig:1D_sweet_spot}B, C -- triangles), whose magnitude and location varies with the slope and midpoint of the correlation function (compare Fig. \ref{fig:1D_sweet_spot}B, C with Fig. \ref{fig:1D_distance_coding}G, H). Looking across different combinations of $\lambda$ and $\sigma_\lambda$, we find a trade-off between the maximum value of the DoC (Fig. \ref{fig:1D_sweet_spot}D) and the distance at which the maximum is realized (Fig. \ref{fig:1D_sweet_spot}E). 

\begin{figure}
    \centering
    \includegraphics[width=0.925\linewidth]{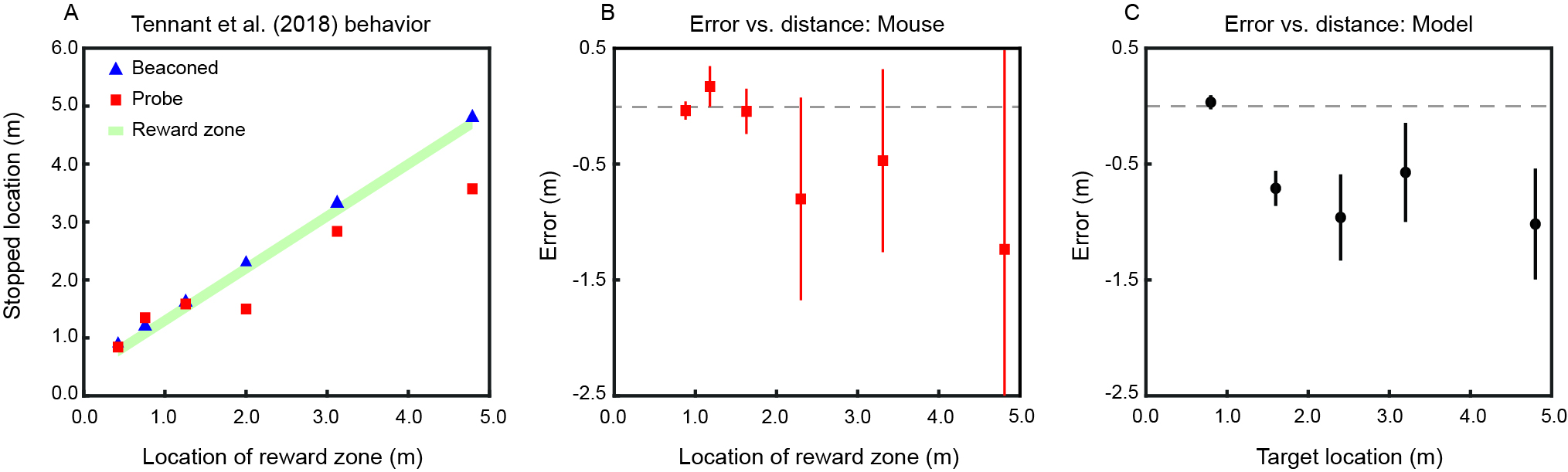}
    \caption{\small \textbf{Experimental data and simulations of heterogeneous grid cell populations are consistent with possible sweet spot in distance coding.} (A) Average stopped location of mice on beaconed and probed trials from Tennant et al.'s (2018) behavioral experiments \cite{tennant2018stellate}. Subpanel is analogous to their Fig. 3D. (B) Error in stopped location, with respect to the true location of the reward zone, of mice from Tennant et al.'s behavioral experiments. Error bars are $\pm$ standard deviation across $4-7$ mice (number of animals differs across each distance). (C) Error in decoded location, with respect to target location, from simulated noisy grid cell populations. $\lambda = 0.80$ m, $\sigma_\lambda = 0.05$ m, and $N = 64$. Error bars are $\pm$ standard deviation across $7$ uniquely sampled populations. }
    \label{fig:distance_decoding}
\end{figure}

The existence of a sweet spot provides a non-intuitive prediction that the accuracy in decoding distance from grid cell activity should be non-monotonic. That is, the error should increase, then decrease, then increase again. We tested this by analyzing previous rodent behavioral experiments. Tennant et al. (2018) trained mice to run along a 1D virtual environment and stop at specific distances, within a $0.2$ m zone, to get a reward \cite{tennant2018stellate}. These distances ranged from $0.88$ m to $4.81$ m. In the case where there was a visual stimulus that marked the start of the reward zone (referred to as ``beaconed'' trials), the mice performed with high accuracy (Fig. \ref{fig:distance_decoding}A -- blue triangles), demonstrating that they were able to use visual cues to guide their navigation. On trials where there was no visual stimulus and the mice had to rely solely on an internal sense of how far they had traveled (referred to as ``probe'' trials), they performed with moderate accuracy, on average stopping short of the reward zone (Fig. \ref{fig:distance_decoding}A -- red squares). Analyzing the publicly available data, we find that the absolute error in the stopped location, with respect to the true reward location, was non-monotonic, increasing when the reward zone location was at $2.3$ m, decreasing when the reward zone was at $3.3$ m, and then increasing again when the reward zone was at $4.8$ m (Fig. \ref{fig:distance_decoding}B). We observe a similar pattern when looking at the error, as a function of distance, on ``non-beaconed'' trials (Fig. \ref{fig:statistical_analysis_distance_decoding_data}A). These trials, like the probe trials, do not have a visual stimulus present, although the mice are able to receive a reward, which is not present on probe trials (see Tennant et al. (2018) for more details on the distinction between probe and non-beaconed trials). The limited number of mice (7 mice total, of which 4--7 completed probe and non-beaconed trials of each target distance) and trials (1 probe trial per experiment, 2 non-beaconed trials per experiment) makes it difficult to assess the statistical significance of this pattern. Indeed, when directly computing the statistics between the errors on probe trials with target distance $2.3$ m and $3.3$ m, we do not find a significant difference (two-sample KS test: $p = 0.37$). To attempt to address this limitation, we perform a bootstrap analysis, where we generate $50$ samples, with replacement, of errors for the two trial types, and compute the average error. We find that the absolute error is significantly larger for the distribution of trials with target distance 2.3 m, as compared to the distribution of trials with target distance 3.3 m (Fig. \ref{fig:statistical_analysis_distance_decoding_data}B; median bootstrapped error on 2.3 m probe trials: $-0.82$ m;  median bootstrapped error on 3.3 m probe trials: $-0.40$ m). However, we stress that this result can only be said to be broadly consistent with the existence of a sweet-spot. More experimental work is necessary before such a claim can be fully supported.

To demonstrate that the behavioral data collected by Tennant et al. (2018) could be explained by a distance coding scheme that leverages heterogeneity in grid properties, we perform a decoding analysis on simulated noisy grid cell populations (see Sec. \ref{subsection: decoding analysis}, for more details). We find that, setting $\lambda = 0.8$ m and $\sigma_\lambda = 0.05$ m [values found in an individual grid module analyzed by Redman et al. (2025)], a similar trend to that observed in the Tennant et al. (2018) emerges (compare Fig. \ref{fig:distance_decoding}C with Fig. \ref{fig:distance_decoding}B). The decoding error is approximately $0.0$ m when the ``target location'' is at the first grid spacing ($x = 1 \cdot \lambda =  0.8$ m). When the target location is at three grid spacings ($x = 3 \cdot \lambda = 2.4$ m), the decoding error is around $-1.0$ m. When the target location is increased to four grid spacings ($x = 4 \cdot \lambda = 3.2$ m), the decoding error improves to approximately $-0.5$ m. By six grid spacings ($x = 6 \cdot \lambda = 4.8$ m), the decoding error becomes nearly $-1.5$ m. Other choices of $\lambda$ and $\sigma_\lambda$ can also generate such patterns (Fig. \ref{fig:distance_decoding_vs_grid_spacing}A, B), but not all combinations lead to error vs. distance relationships that are consistent with the data. This makes specific predictions about the grid properties that Tennant et al. would have found had they recorded from MEC. 

\subsection{A thresholded rate-based model of distance coding in 1D}
\label{subsection: Thresholded ratebased model}
The distance by de-correlation code we have thus far studied has two assumptions. First, the monotonic relationship between population vector correlation and distance occurs at integer multiples of the underlying grid spacing. And second, the current grid cell activity must be able to be compared to the grid cell activity at an initial location (e.g., $x = 0$). Whether and how these assumptions can be realized in a biologically plausible manner is an important question. We therefore develop a thresholded rate-based model in which both of these can be addressed. For simplicity, we again restrict ourselves to the 1D setting.

\begin{figure}
    \centering
    \includegraphics[width=0.75\linewidth]{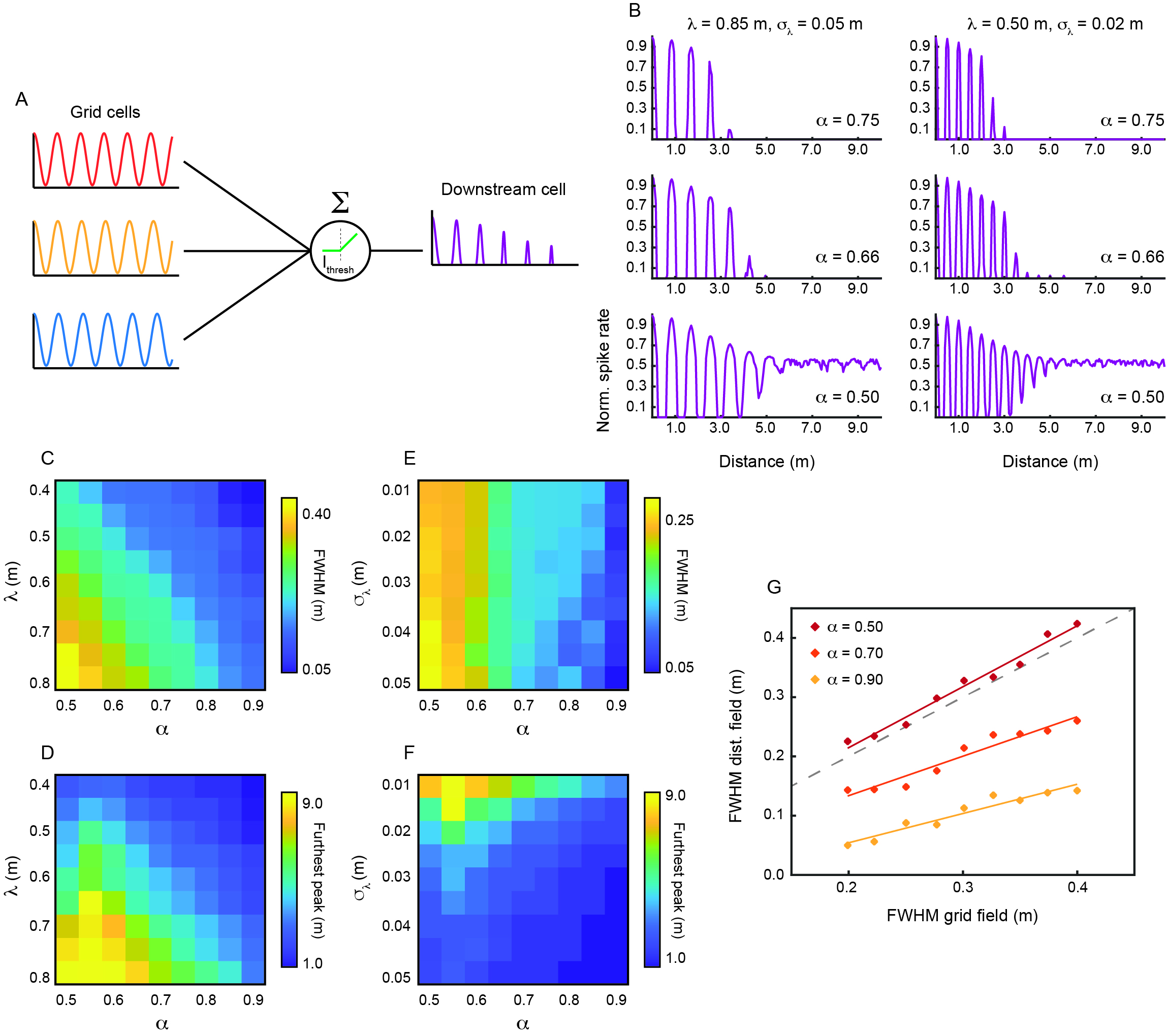}
    \caption{\small \textbf{A thresholded rate-based model enables distance by de-correlation coding in 1D.} (A) Schematic of a downstream neuron that receives input from $N$ grid cells with variable grid spacing. This downstream neuron's firing rate is determined by a thresholded sum of the inputs, with anything less than the $I_\text{thresh}$ being subthreshold and not leading to a response. (B) Example firing rates, as a function of distance, for downstream neurons modeled using different choices of $\lambda$, $\sigma$, and $\alpha$. (C) Average FWHM of peaks in firing rate, as a function of distance, of downstream neuron, for different choices of $\lambda$ and $\alpha$. (D) Average location of  furthest peak in firing rate of downstream neuron, for different choices of $\lambda$ and $\alpha$. (C)--(D) $\sigma_\lambda = 0.02$ m. (E)--(F) Same as (C)--(D), but with $\lambda = 0.5$ m and $\sigma_\lambda$ being varied. (G) FWHM of the downstream neuron's firing peaks, as a function of the FWHM of the grid firing fields. $\sigma_\lambda = 0.02$ m. (B)--(G) $N = 128$ and $25$ independent populations.}
    \label{fig:1D-rate_based_distance_model}
\end{figure}

We consider a downstream neuron receiving input from $N$ grid cells from one module that have grid spacings sampled from $\mathcal{N}(\lambda, \sigma_\lambda^2)$ and are locked to $\phi = 0$ (Fig. \ref{fig:1D-rate_based_distance_model}A). This downstream neuron has the following firing rate,
\begin{equation}
    H(x) = \sigma \left( \sum_{i = 1}^N G_{\lambda_i, \omega}(x) - I_\text{thresh} \right),
\end{equation}
where $\sigma(\cdot)$ is a rectified linear activation function, with $\sigma(g) = g$ if $g > 0$ and $g = 0$ otherwise. Anything less than $I_\text{thresh}$ is subthreshold and does not lead to firing, whereas anything greater than $I_\text{thresh}$ is suprathreshold and leads to a linear firing rate. We can imagine scaling $I_\text{thresh} = \alpha G^{\max} N$, where $G^{\max}$ is the average peak firing rate of grid cells at the grid field centers and $\alpha \in [0, 1]$. As $\alpha$ increases, the downstream neuron becomes more selective, firing only when a larger fraction of the grid cells are firing near their peak rate.

When $\sigma_\lambda > 0$, the $N$ grid cells will de-correlate with distance, as demonstrated in Fig. \ref{fig:1D_distance_coding}. This means that $\sum_i G_{\lambda_i, \omega}(n \cdot \lambda)$ will monotonically decrease as $n$ increases, for $n \in \mathbb{Z}$. This de-correlation also comes with an increase in the total grid firing at non-integer multiples of the grid spacing (e.g., $G_{\lambda_i, \omega}[(n + 1/2) \cdot \lambda]$ increases as $n$ increases). However, we may expect that setting $I_\text{thresh}$ sufficiently large will repress the response of the downstream neuron at non-integer multiples of the grid spacing (Fig. \ref{fig:1D-rate_based_distance_model}A).

Simulating this model for $\lambda = 0.85$ m and $\sigma_\lambda = 0.05$ m, we find that, $\alpha = 0.75$ leads to several sharp peaks in firing rate of the downstream neuron, as a function of distance (Fig. \ref{fig:1D-rate_based_distance_model}B, top left). The amplitude of these peaks decreases with distance, up until approximately $3.5$ m, after which the downstream neuron is silent. Decreasing $\alpha$ leads to wider peaks that persist at further distances (Fig. \ref{fig:1D-rate_based_distance_model}B, middle and bottom left). When $\alpha = 0.5$, the monotonic decrease of the downstream neurons firing rate, at integer multiples of the grid spacing, persists up until $5$ m, after which the firing rate plateaus to a constant baseline rate. We find similar results for a population of grid cells with $\lambda = 0.50$ m and $\sigma_\lambda = 0.02$ m (Fig. \ref{fig:1D-rate_based_distance_model}B, right).

To explore the trade-off between selectivity of the downstream neuron's response, as a function of distance, and its ability to represent a range of distance, we perform sweeps across the thresholded rate-based models, sampling populations of grid cells with varying $\lambda$ and $\sigma_\lambda$ and different choices of $\alpha$. For each resulting model, we measure the FWHM of the peaks in the firing rate of the downstream neuron and the location of its most distant peak. Consistent with our initial simulations, we find that -- for a fixed $\lambda$ and $\sigma_\lambda$ -- increasing $\alpha$ leads to a smaller peak FWHM (Fig. \ref{fig:1D-rate_based_distance_model}C, E), corresponding to greater selectivity, and smaller furthest peak location (Fig. \ref{fig:1D-rate_based_distance_model}D, F), corresponding to a reduced range of distance coding.

Given that decreasing $\alpha$ leads to increased FWHM of the downstream neuron's peak firing rate, but increases the range over which distance can be coded, we wondered how the FWHM of peaks in the downstream neuron's firing rate compared to the FWHM of the grid fields. In particular, if $\alpha$ is such that the FWHM of the downstream neuron is smaller than the FWHM of the grid cells, then the downstream neuron has gained some spatial selectivity. We find that, for $\alpha = 0.7$ and $\alpha = 0.9$, the FWHM of the downstream neuron is well under the FWHM of the grid fields (Fig. \ref{fig:1D-rate_based_distance_model}G). In particular, when $\alpha = 0.9$, the FWHM of the grid cells are $\approx 3\times$ greater than the FWHM of the downstream neuron. When $\alpha = 0.5$, the FWHMs are approximately equal (Fig. \ref{fig:1D-rate_based_distance_model}G).

Collectively, our thresholded rate-based model -- while simple -- demonstrates that distance can be encoded by the firing rate of downstream neurons in a simple manner, with a choices of threshold controlling specificity and range.

\subsection{Grid variability and distance coding in 2D}
\label{subsection: Grid variability 2D}

We extend our theory and numerical analysis to 2D environments. For sake of brevity, we present the theory in Sec. \ref{subsection: Analytic 2D extension of theory}. Moving to 2D introduces two changes to our numerical simulations. First, unlike 1D environments where grid cells can be locked to the starting location \cite{campbell2021distance}, in 2D environments grid cells are found to have distributed grid phase \cite{hafting2005microstructure}. Therefore, for each noisy grid cell we simulate, we sample its grid phase from a uniform distribution over all possible translations. And second, we remove the assumption that the grid orientation is set to $0^\circ$. Therefore, we consider the existence of non-zero variability in grid orientation ($\sigma_\theta > 0^\circ$). In addition, when simulating grid cells coming from multiple modules, we assume that each module has its own orientation \cite{stensola2012entorhinal}.

\begin{figure}
    \centering
    \includegraphics[width=0.825\linewidth]{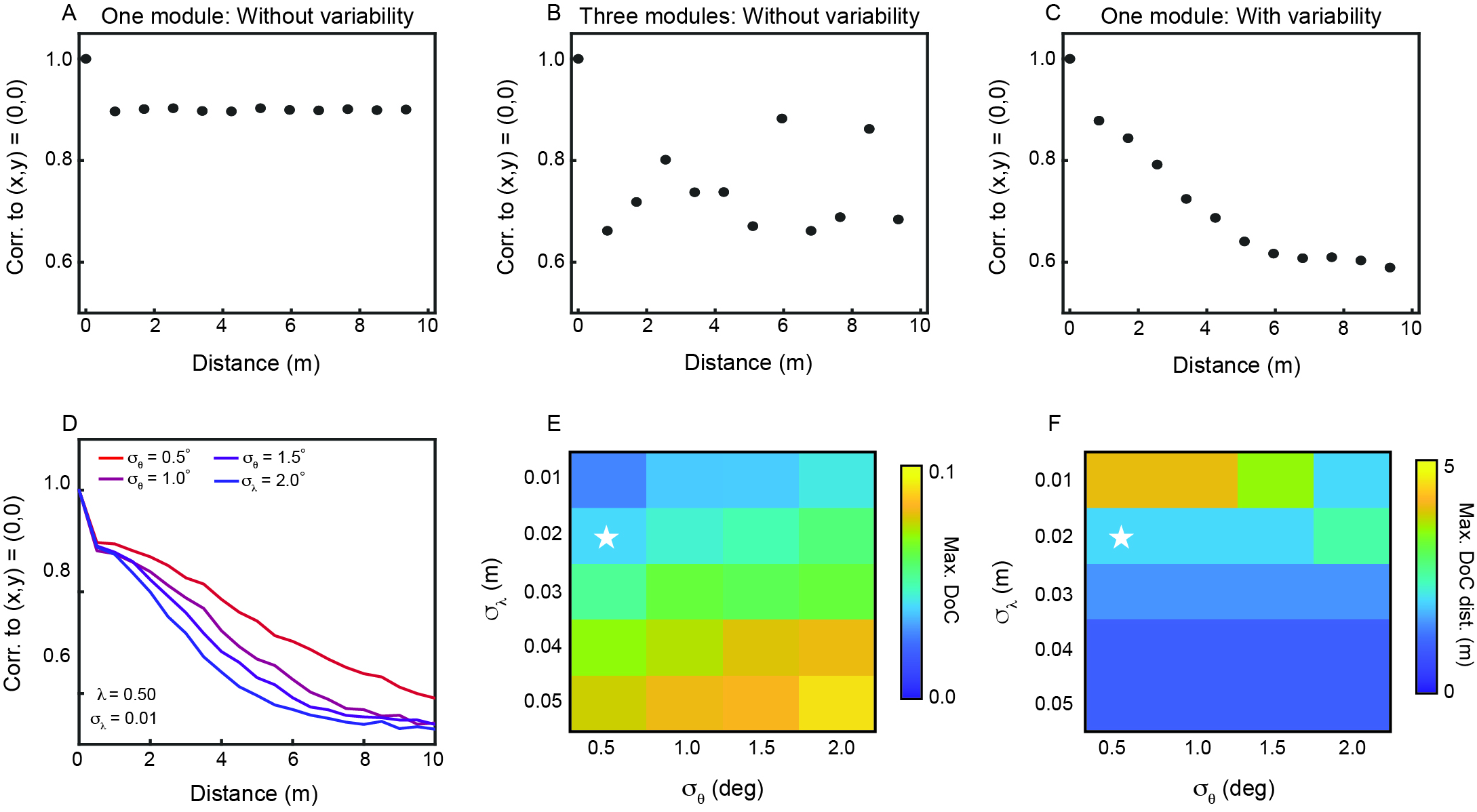}
    \caption{\small \textbf{Variability in grid properties leads to monotonic relationship between population vector correlation and distance in 2D, along the $x-$axis.} (A)--(C) Correlation of population vectors, corresponding to distances at integer multiples of grid spacing along the $x$--axis, with the population vector corresponding to the center of the 2D arena, for different populations of grid cells. Same choices of parameters as used in Fig. \ref{fig:1D_distance_coding}D--F, with the heterogenous grid module having $\sigma_\theta = 5^\circ$. (D) Population vector correlation, as a function of distance, for different grid orientation variabilities, $\sigma_\theta$. $\lambda$ is fixed at $0.50$ m and $\sigma_\lambda$ is fixed at $0.01$ m. (E) Maximum DoC value for different combinations of $\sigma_\lambda$ and $\sigma_\theta$. (F) Distance at which the maximum DoC value was achieved, for different combinations of $\sigma_\lambda$ and $\sigma_\theta$. (E)--(F) White star denotes previously found value of $\sigma_\lambda$ and $\sigma_\theta$ in analysis of experimentally recorded grid cells \cite{redman2025robust}. For all plots, $\theta = 0^\circ$, $N = 128$, and $25$ unique populations were simulated.}
    \label{fig:2D_distance_coding}
\end{figure}

We start by considering distance along the $x-$axis. That is, we restrict ourselves to a 1D slice through the 2D space. We find that a population of grid cells from a single grid module, with a fixed grid spacing and orientation, does not show any distance dependent de-correlation with the population vector at the center of the environment, $(x, y) = (0, 0)$ (Fig. \ref{fig:2D_distance_coding}A). Similarly, we find that a population of grid cells coming from three modules with fixed grid properties exhibits a non-monotonic relationship between distance and correlation (Fig. \ref{fig:2D_distance_coding}B). However, a population of grid cells from a single grid module with heterogeneity in grid properties does have a monotonic decay of correlation, as a function of distance (Fig. \ref{fig:2D_distance_coding}C). While the relationship is similar to the 1D case (where $\sigma_\theta = 0^\circ$ and there is no distribution of grid phase -- Fig. \ref{fig:1D_distance_coding}F), there is an important difference. In particular, the asymptotic correlation value is $\approx 0.6$, instead of $\approx 0.8$. This implies that there is greater de-correlation in 2D environments, with $\sigma_\theta$ and the distribution of phases acting to further separate population vectors. This can be seen clearly when we fix the $\lambda$, $\theta$, and $\sigma_\lambda$, and vary $\sigma_\theta$ (Fig. \ref{fig:2D_distance_coding}D). In addition, we find that greater $\sigma_\theta$ leads to faster de-correlation (Fig. \ref{fig:2D_distance_coding}D), as was found with increasing grid spacing variability in the 1D setting (Fig. \ref{fig:1D_distance_coding}G). 

By shaping the distance dependence of the population vector correlation, $\sigma_\theta$ also shapes the trade-off between the distinguishability of different locations and the distance over which the de-correlation occurs. In particular, for a given $\lambda$ and $\theta$, we find that increasing either $\sigma_\lambda$ or $\sigma_\theta$ leads to an increase in maximum DoC (Fig. \ref{fig:2D_distance_coding}E), while at the same time leading to a rapid decrease in location of maximum DoC (Fig. \ref{fig:2D_distance_coding}F). We find that $\sigma_\theta = 0.5^\circ$ and $\sigma_\lambda = 0.02$ m, which was the average variability found across all experimentally recorded modules previously analyzed \cite{redman2025robust}, achieves both an intermediate value of maximum DoC and maximum DoC location (Fig. \ref{fig:2D_distance_coding}E, F). This suggests that grid property variability may be tuned to enable accurate of distance coding, over a non-trivial range of distance.  

\begin{figure}
    \centering
    \includegraphics[width=0.825\linewidth]{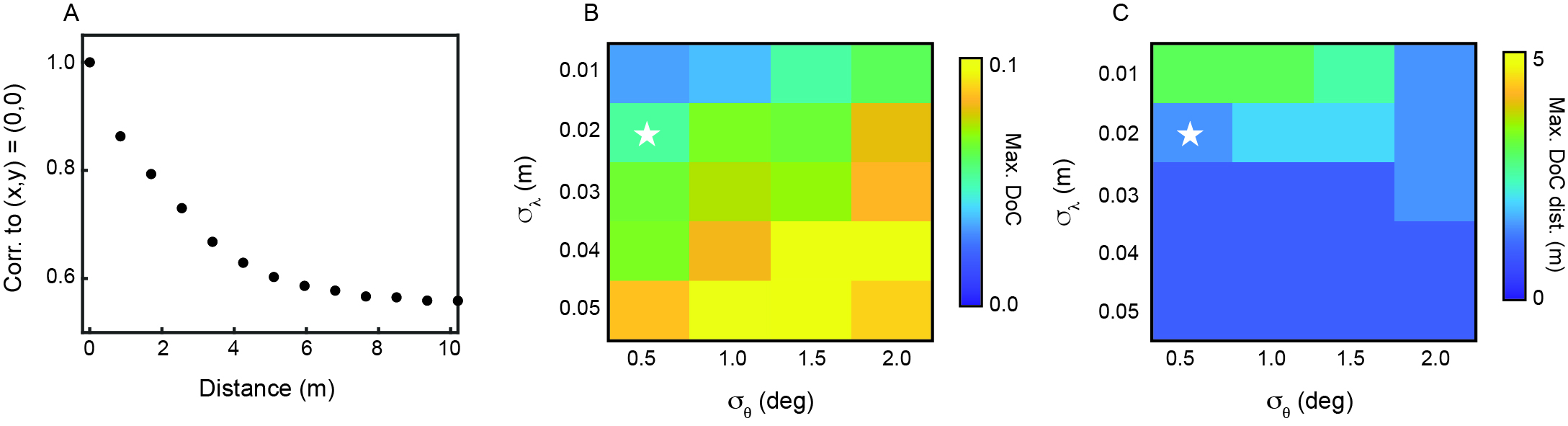}
    \caption{\small \textbf{Distance by de-correlation coding exists in 2D when considering all distances.} (A) Correlation of population vectors, corresponding to distances at integer multiples of grid spacing with the population vector corresponding to the center of the 2D arena, for heterogeneous grid cells. Unlike Fig. \ref{fig:2D_distance_coding}, not just positions along the $x-$axis are considered. (B) Maximum DoC value for different combinations of $\sigma_\lambda$ and $\sigma_\theta$. (C) Distance at which the maximum DoC value was achieved, for different combinations of $\sigma_\lambda$ and $\sigma_\theta$. (B)--(C) White star denotes previously found value of $\sigma_\lambda$ and $\sigma_\theta$ in analysis of experimentally recorded grid cells \cite{redman2025robust}. For all plots, same parameters as Fig. \ref{fig:2D_distance_coding}.}
    \label{fig:2D_distance_coding_all_distance}
\end{figure}

To understand whether a distance by de-correlation grid code could be effective in general in 2D space, we consider a ``base'' grid cell [with grid phase $\phi = (0, 0)$] and find the location of all of its grid fields. We then compute the population vector correlation (across a heterogeneous grid cell population) at these locations with the center of the environment  $(x, y) = (0, 0)$. We then bin the distance of all of these points into integer multiples of the grid spacing, taking the average population vector correlation in each bin. We again find strong monotonic behavior of the population vector correlation as a function of distance (Fig. \ref{fig:2D_distance_coding_all_distance}A). In addition, performing the same sweeps over $\sigma_\lambda$ and $\sigma_\theta$, we again observe a trade-off between maximum DoC and its location (Fig. \ref{fig:2D_distance_coding_all_distance}B, C). The previously experimentally measured values of $\sigma_\lambda$ and $\sigma_\theta$ remains a point that balance this trade-off (Fig. \ref{fig:2D_distance_coding_all_distance}B, C). Both of these results, monotonic decay of correlation with 2D distance and increased rate of decay with greater $\sigma_\lambda$ or $\sigma_\theta$, are consistent with explicit predictions from the 2D theory (Sec. \ref{subsection: Analytic 2D extension of theory}). 

\section{Discussion}

The organization of grid cells into discrete modules \cite{stensola2012entorhinal} has led many theoretical investigations on their computational role to make the corollary assumption that, within a module, all grid cells are identical, up to translation \cite{fiete2008grid, burak2009accurate, sreenivasan2011grid, mathis2012optimal, wei2015principle, stemmler2015connecting}. While experimental investigation has found evidence that grid cells are involved in the computing of distance between points \cite{jacob2017medial, tennant2018stellate, duncan2025grid}, supporting their putative role in spatial navigation, the requirement of a single grid spacing and orientation per grid module  -- which is assumed by the theoretical work that underlies the field's understanding of the computational role of grid -- has recently been challenged \cite{redman2025robust}. In particular, analysis of state-of-the-art electrophysiological recordings of medial entorhinal cortex \cite{gardner2022toroidal} identified small, but robust heterogeneity in grid properties within the same grid module \cite{redman2025robust}. This variability leads to a breaking of the grid cell population activity translation symmetry and suggests that individual grid modules may have considerably more computational capacity than was previously assumed. However, what computational properties a heterogeneous grid module has remains unclear. This is important to address, as the variability in grid properties implies that, \textit{a priori}, the previously developed grid cell theory may not hold. 

In this work, we hypothesize that heterogeneity of grid properties, within a grid module, generates monotonic distance dependent de-correlation of population activity (Fig. \ref{fig:schematic}), enabling simple decoding of distance. Analyzing populations of distance-tuned MEC neurons recorded in mice running along 1D environments, we find that such a decay of population correlation with distance can exist (Fig. \ref{fig:1D_distance_coding_Campbell_data}B, D). We develop a theory for the effect heterogeneity in grid spacing has on population vector correlation, as a function of distance, in one-dimension, which predicts a trade-off between the range over which the de-correlation occurs (maximized by increasing $\lambda$ or decreasing $\sigma_\lambda$) and the distinguishability between different locations (maximized by decreasing $\lambda$ or increasing $\sigma_\lambda$). Numerical simulations support the presence of such a trade-off in both 1D (Fig. \ref{fig:1D_distance_coding}) and 2D (Fig. \ref{fig:2D_distance_coding_all_distance}), where variability in grid orientation further influences how the de-correlation occurs. These results provide new insight on why the observed range of grid properties, within a given module, is relatively limited. In particular, encoding distance requires balancing both range and distinguishability, constraining $\sigma_\lambda$ and $\sigma_\theta$ towards small (but not too small) values. Such small variability is not optimal for encoding local spatial position, the previously suggested role for grid heterogeneity \cite{redman2025robust}, as increasing variability only improves accuracy. This distance information at integer multiples of the grid spacing can be encoded by the firing rate downstream neurons that receive inputs from grid cells with similar grid phase and that apply a threshold based on the percent of grid cells that should be maximally active (Fig. \ref{fig:1D-rate_based_distance_model}). Different thresholds lead to different trade-offs between specificity and range of distances that can be encoded.

Our simulations led us to identify that a distance by de-correlation code has a ``sweet spot'', at which a further distance is better encoded than some nearer distances (Fig. \ref{fig:1D_sweet_spot}). We find preliminary evidence for this in previously published rodent behavioral experiments \cite{tennant2018stellate} and find that decoding distance from the population vector correlation of noisy grid cells generates results that are consistent with these experiments (Fig. \ref{fig:distance_decoding}). This is a non-trivial prediction, as distortions to the periodic order of grid cells \cite{krupic2015grid, ginosar2023grid}, through the presence of sheering or compression, would be expected to lead only to an increase in distance coding error. This is also true of errors arising from path integration \cite{stangl2020sources}, which would compound with distance. However, given the small number of mice, small number of trials, possibility of de-motivation across the experiment, and possible changes to the latent states of the mice, more work must be done before a full conclusion can be made on the existence of a ``sweet spot''.

Path integration has been another, closely related, computational role proposed for grid cells \cite{mcnaughton2006path, burak2009accurate} which has seen experimental support \cite{gil2018impaired, clark2024task}. Continuous attractor networks (CANs) are one class of models that could enable path integration via grid cells \cite{burak2009accurate} and extensive experimental work has been performed testing aspects of CANs in MEC data \cite{trettel2019grid, gardner2019correlation, gardner2022toroidal}. An important question then is, how does variability in grid spacing and orientation affect CANs? Or more precisely, can CANs exist when the underlying grid cells are heterogeneous? We believe there are three reasons to suspect that the answer may be yes. First, the MEC data that was analyzed and found to have grid property variability was one of the datasets where a signature of CANs, namely the population activity lying along a toroidal manifold, was found \cite{gardner2022toroidal}. Thus, at least this specific signature of a CAN can co-exist with heterogeneous grid cells. Second, recurrent neural networks (RNNs) trained to perform path integration have been found to develop hidden units with grid-like tuning \cite{banino2018vector, cueva2018emergence, sorscher2023unified}. Furthermore, it has been argued that the RNNs develop structure that is consistent with CANs \cite{sorscher2023unified}. Redman et al. (2025) applied the same analysis to the RNN grid-like units as they did to the MEC neural data, and found that -- like the recorded grid cells -- grid-like units in path integrating RNNs exhibited variability in their grid spacing and orientation \cite{redman2025robust}. Thus, not only can grid heterogeneity co-exist with signatures of CANs, it can co-exist with path integration, and the variability may even have a normative explanation. And third, ring attractors -- one-dimensional CANs that have been found to underlie the head-direction circuit in \textit{Drosophilia} \cite{kim2017ring} and have strong experimental support to exist in rodents \cite{chaudhuri2019intrinsic} --  typically assume perfect Gaussian receptive fields. This assumption is not met in experimentally recorded data. However, recent theoretical work has shown it is possible to nevertheless develop a network with random, disordered structure that -- in the limit -- becomes a true CAN \cite{clark2025symmetries}. Thus, at least in one-dimension, some degree of heterogeneity can be exactly supported by CANs.

Recent work by Chaudhuri et al. (2025) similarly examines the correlation of grid cell activity as a function of distance \cite{chaudhuri2025not}. However, the conclusion reached is that the population vector of grid cell activity is ``scrambled'' (or orthogonalized) at distances beyond the width of a grid field. On the face of it, this appears to contradict our theory. We believe these two views [the one presented here and the one presented in Chaudhuri et al. (2025)]  can be reconciled in the following ways. First, like previous theoretical work, Chaudhuri et al. (2025) assume that a single grid spacing and orientation exists in each model is made. This leads the authors to primarily consider the activity across multiple grid modules. But, when considering populations of grid cells from the same grid module, Chaudhuri et al. (2025) find that grid cell recordings done in 1-D \cite{campbell2021distance} exhibit periodic correlations, as a function of distance, that appear in some mice to decay monotonically [Fig. 3b in Chaudhuri et al. (2025)]. In addition, Chaudhuri et al. (2025) find that grid cell recordings done in 2-D \cite{gardner2022toroidal} exhibit periodic correlations, as a function of distance, that appear to decay [Fig. 2c in Chaudhuri et al. (2025)], when considering populations of grid cells from individual grid modules. Inspired by these observations, we analyzed the same 1D dataset and found that the periodic correlations that decayed with distance were robust to using cosine similarity,  so as to avoid trivial decay (Fig. \ref{fig:1D_distance_coding_Campbell_data}). Thus, the results of Chaudhuri et al. (2025) are not inconsistent with our theory, when activity from a single grid module is examined. And second, one of the major hypotheses of Chaudhuri et al. (2025) is that integrating grid cell activity across multiple grid cell modules enables a more complete orthogonalization to occur. Because our theory is restricted to individual grid modules with variability in grid spacing and orientation, our work is not in conflict with this idea. Thus, we believe that our theory is in fact complimentary to that developed by Chaudhuri et al. (2025): grid cell populations across grid modules have activity that gets scrambled, enabling efficient separation of nearby locations, but grid cell populations within a single heterogeneous grid module maintain structure that enables a more global distance metric. 

\textbf{Limitations.} The distance by de-correlation code we study enables coding of distance only at integer multiples of the grid spacing. While this can provide information that is behaviorally relevant (e.g., is a rodent's nest near-by, $< 1$ m, or far away, $> 5$ m), it does not provide finer-scale distance information. However, we note that in some cases, there may be an obvious spacing that is most natural to compute distance with respect to (e.g., distance between aisles in a grocery store). This can also be true in ``abstract'' spaces, where grid cells have been known to emerge \cite{constantinescu2016organizing, aronov2017mapping}. For instance, in recent work where non-human primates ``traveled'' between different images, there was a set spacing between each image \cite{neupane2024mental}, providing a natural reference distance. While our thresholded rate-based model provides a simple mechanism for encoding distance information at integer multiples of the grid spacing, more biologically plausible approaches need to be considered.

The theory we develop assumes no noise. This limits its applicability to real grid cell recordings, where grid cell firing fields often exhibit significant departure from pure Gaussians, and can even exhibit grid field specific firing rates \cite{diehl2017grid, dunn2017grid, ismakov2017grid}. However, we find that simulating populations of grid cells, modeled as Poisson processes, generate results that are consistent with the theory. This suggests it is capable of providing insight that is relevant, even in more realistic settings. 

While our decoding analysis provides results that are consistent with Tennant et al.'s (2018) experiments, there is an important difference. Decoding using the population vector correlation leads to an increase in error at a distance equal to twice the grid spacing ($1.6$ m) (Fig. \ref{fig:1D_distance_coding}C). This contrasts the experimental finding that accuracy stays high up to approximately $1.5$ m (Fig. \ref{fig:1D_distance_coding}B). In addition, the behavioral experiments were only performed with, at most, 7 mice \cite{tennant2018stellate}, making it challenging to assert claims of statistically significance to the apparent sweet spot. Therefore, it remains to be seen whether the sweet spot is due to the specifics of the experiment or is instead a more general property. This can be tested in future experiments. Similarly, the existence of decaying population vector correlation with distance was seen in some of the recordings from Campbell et al. (2021) (Fig. \ref{fig:1D_distance_coding_Campbell_data}). However, this data was recorded in mice as they traveled hundreds of meters. These extensive distances likely reduced the periodicity in the autocorrelations. This weakened the amount of structure that could be exploited by a distance by de-correlation code. In addition, the identity of the grid module each distance-tuned neuron belonged to was not provided, which led us to make \textit{ad hoc} decisions of grouping. Future experiments could perform more extensive recordings on distance-tuned MEC neurons over shorter intervals (e.g., $10$s of meters per trial), with anatomical and functional based clustering of cells in modules.

Finally, we have assumed that grid cells encode distance through their joint population activity. This is consistent with many theoretical works, but experimental work has found evidence that grid cells can encode distance and time through distance and/or temporally selective firing \cite{kraus2015during}. In addition, non-grid cells in MEC have shown firing rates that increase (``ramp-like'' responses) with distance and/or time \cite{neupane2024mental}. Thus, future work is needed to understand if and whether grid cells are necessary for encoding distance. However, we note that the rate-based changes in grid cells found by Kraus et al. (2015) occured when the rodent ran on a treadmill \cite{kraus2015during}, a somewhat unnatural behavior that also led to grid cells losing their periodic responses. Furthermore, the ramp-like responses found by Neupane et al. (2024) in MEC were found encode temporal ranges up to 3 seconds \cite{neupane2024mental}. It may be that a ramp-like code would not be possible over as large a range of distance as was found by our distance by de-correlation code, although this remains to be tested.

\textbf{Future directions.} While detailed analysis has shown the presence of robust within module grid property variability \cite{redman2025robust}, broader characterization of such heterogeneity remains to be done. Performing this analysis on more datasets, across different model species (e.g., mouse, rat, bat) will not only increase our understanding of the prevalence of grid heterogeneity, but will also provide greater understanding of how the trade-off between the range over which de-correlation occurs and the distinguishability between distances is balanced. 

Computational advantages of having grid cells with variable properties within a single grid module, beyond distance coding, may exist. Indeed, it is well appreciated in the computational neuroscience community that heterogeneity in neuron responses can be useful for increasing robustness \cite{shamir2006implications, chelaru2008efficient, perez2021neural}. Future work should explore whether other uses of grid heterogeneity exist. In addition, how the heterogeneity arises, and how it is maintained by a 2D attractor network, remains an outstanding open question that should be addressed.

\textbf{Outlook.} It is always possible to find aspects of a computational model that are not fully consistent with experimental results. The motivation of this work is not to find fault with previous theoretical work, but to demonstrate that weakening an assumption made by this theory -- all grid cells in the same module are identical, up to translation -- greatly expands the computational capacity of individual grid modules. In addition, our results -- along with the results of Redman et al. (2025) \cite{redman2025robust} -- illustrate that many of the ideas that exist with prior theoretical work (e.g., the ability to decode local spatial positions, the ability to decode distances) are still possible with a heterogeneous population of grid cells. We hope that future work will continue to integrate these two perspectives and shed greater light on the extent to which populations of grid cells with variability in grid spacing and orientation can support the important cognitive processes that grid cells are believed to underlie. Finally, we highlight the fact that this work motivates the question, why are there multiple grid modules? We hope that by showing that a single grid module with small, but robust variability in grid spacing and orientation can perform computations that have been assumed to require multiple grid modules, we can further hone in on what exactly multiple grid modules enable. 


\section{Materials and Methods}
\subsection{Analysis of 1-D grid cell data}
\label{subsection: analysis of 1d grid cell data}

We analyze the Neuropixel \cite{jun2017fully} MEC recordings from Campbell et al. (2021) using the publicly available dataset\footnote{\texttt{https://plus.figshare.com/articles/dataset/VR\_Data\_Neuropixel\_supporting\_Distance-tuned\_neurons\_drive\\\_specialized\_path\_integration\_calculations\_in\_medial\_entorhinal\_cortex\_/15041316}} and code\footnote{\texttt{https://zenodo.org/records/5138030}}. Because the autocorrelation was found to decay with distance (Fig. \ref{fig:1D_distance_coding_Campbell_data}A, C), just computing the correlation between the population activity at $x = 0$ and at $x > 0$, we would expect to see a decrease. Therefore, to avoid observing a trivial distance dependence on correlation, we normalize the population activity at each position by its norm. That is, we compute 
\begin{equation}
\label{eq: normalized grid cell correlation}
    C(x) = \frac{G(x)^T G(0)}{||G(x)||_2 \cdot ||G(0)||_2},
\end{equation}
where $G(x)$ is the grid cell population activity vector at distance $x$. Eq. \ref{eq: normalized grid cell correlation} is equivalent to the cosine similarity. 

Having computed the correlation function, $C(x)$, we plot its peaks. This is because, unlike the simulations where we know exactly what the average grid spacing is and can plot $C(x)$ at integer multiples of it, it is not clear what the grid spacing of the recorded grid cells are. In addition, because the autocorrelation decays with distance, the grid spacing may change slightly with distance. Therefore, to be most conservative, we plot the peaks of $C(x)$. 

\subsection{Derivation of the population vector correlation}
\label{subsection: Exact derivation of population vector correlation}

We begin with the power spectrum from Eq. \ref{eq: expected power distribution} and we combine the exponentials for each peak $k$:
\begin{equation}
    P_k(\xi) = \exp\left(-\frac{(\xi - \mu_k)^2}{\sigma_k^2}\right) e^{-\omega^2\xi^2} = \exp\left[ -\left( \frac{1}{\sigma_k^2} + \omega^2 \right)\xi^2 + \left(\frac{2\mu_k}{\sigma_k^2}\right)\xi - \frac{\mu_k^2}{\sigma_k^2} \right].
\end{equation}
We can complete the square by defining an effective variance $\tilde{\sigma}_k^2$ and an effective mean $\tilde{\mu}_k$
\begin{equation}
    \frac{1}{\tilde{\sigma}_k^2} = \frac{1}{\sigma_k^2} + \omega^2 \implies \tilde{\sigma}_k^2 = \frac{\sigma_k^2}{1 + \omega^2\sigma_k^2}.
\end{equation}
\begin{equation}
    \frac{2\tilde{\mu}_k}{\tilde{\sigma}_k^2} = \frac{2\mu_k}{\sigma_k^2} \implies \tilde{\mu}_k = \frac{\mu_k}{1 + \omega^2\sigma_k^2}.
\end{equation}
Substituting these into the exponent and isolating the residual constant term yields
\begin{equation}
    P_k(\xi) = \exp\left(-\frac{\omega^2\mu_k^2}{1+\omega^2\sigma_k^2}\right) \exp\left(-\frac{(\xi - \tilde{\mu}_k)^2}{\tilde{\sigma}_k^2}\right) \equiv \tilde{w}_k \exp\left(-\frac{(\xi - \tilde{\mu}_k)^2}{\tilde{\sigma}_k^2}\right).
\end{equation}

By the Wiener--Khinchin theorem, the autocorrelation $C(x)$ is the Fourier transform of the power spectrum $S(\xi) = \sum_k P_k(\xi)$. Integrating each term using the standard identity $\int e^{-(\xi-\mu)^2/\sigma^2}e^{i\xi x}\,d\xi = \sigma\sqrt{\pi}\,e^{i\mu x}\,e^{-\sigma^2 x^2/4}$, we have
\begin{equation}
    C(x) \propto \sum_{k \in \mathbb Z} \tilde{w}_k \tilde{\sigma}_k \sqrt{\pi} \, e^{i\tilde{\mu}_k x} e^{-\tilde{\sigma}_k^2 x^2/4}
\end{equation}
By symmetry, $\tilde{\mu}_{-k} = -\tilde{\mu}_k$, $\tilde{\sigma}_{-k} = \tilde{\sigma}_k$, and $\tilde{w}_{-k} = \tilde{w}_k$. Summing over all positive and negative integers $k$ gives
\begin{equation}
    C(x) \propto \sqrt{\pi} \sum_{k \in \mathbb Z} \tilde{w}_k \tilde{\sigma}_k e^{-\tilde{\sigma}_k^2 x^2/4} \cos(\tilde{\mu}_k x).
\end{equation}
To get $C(0) = 1$, which is appropriate for the population vector correlation, it is necessary to divide by $C(0) = \sqrt{\pi} \sum_{k} \tilde{w}_k \tilde{\sigma}_k$. This yields the expression:
\begin{equation}
\label{eq: Exact population vecotr correlation}
    C(x) = \frac{1}{\sum_{k} \tilde{w}_k\,\tilde{\sigma}_k} \sum_{k} \tilde{w}_k\,\tilde{\sigma}_k\; e^{-\tilde{\sigma}_k^2 x^2/4}\,\cos(\tilde{\mu}_k x).
\end{equation}

To interpret the analytic form (Eq. \ref{eq: Exact population vecotr correlation}), we examine the behavior of the exponential decay $\exp(-\tilde{\sigma}_k^2 x^2/4)$. Because the spectral width of the $k$-th harmonic scales linearly with $k$ (i.e., $\sigma_k \propto |k|$), the variance $\sigma_k^2$ grows quadratically. Thus, for any distance $x \gtrsim \lambda$, the terms corresponding to large $k$ are suppressed extraordinarily fast, making the sum heavily dominated by the first few harmonics (i.e., small $k$). In addition, since the Gaussian width \(\omega \ll \lambda\) and the variance \(\sigma_\lambda \ll \lambda\), we get
\begin{equation}
    \omega^2\sigma_k^2 = \omega^2 \left( \frac{2\pi k \sigma_\lambda}{\lambda^2} \right)^2 = \left( 2\pi k\right)^2 \left( \frac{\omega \sigma_\lambda}{\lambda^2} \right)^2 \ll 1.
\end{equation}
Under this limit, the effective parameters simplify back to $\tilde{\sigma}_k \approx \sigma_k$, $\tilde{\mu}_k \approx \mu_k$, and $\tilde{w}_k \approx \exp(-\omega^2\mu_k^2) = w_k$. Substituting these into the exact solution recovers the simplified approximation
\begin{equation}
    C(x) \approx \frac{1}{\sum_{k} w_k\,\sigma_k} \sum_{k} w_k\,\sigma_k\; e^{-\sigma_k^2 x^2/4}\,\cos(\mu_k x)
\end{equation}
Evaluating the correlation at \(x=n\lambda\),

\begin{equation}
    \label{eq: Exact correlation function}
    C(n\lambda)\;\approx\;
    \frac{1}{\sum_k w_k\sigma_k} \sum_k w_k\sigma_k\,e^{-\sigma_k^2(n\lambda)^2/4},
\end{equation}
and the decay rate is governed by,
\begin{equation}
    \sigma_k^2(n\lambda)^2/4 = (n\pi k\sigma_\lambda / \lambda)^2
\end{equation}
Eq. \ref{eq: Exact correlation function} gives us the three predictions for $C(n \lambda)$ presented in Sec. \ref{subsection: Grid variability 1D} with the simplified field envelope:
\begin{itemize}
  \item \textbf{(P1)} Monotonic decay with increasing $n$, as $C(n\lambda) \propto e^{-n^2}$;
  \item \textbf{(P2)} Faster decay with greater variability in grid spacing, as $C(n\lambda) \propto e^{-{\sigma_\lambda}^2}$;
  \item \textbf{(P3)} Slower decay for larger $\lambda$, as $C(n\lambda) \propto e^{-1/\lambda^2}$.
\end{itemize}

\subsection{Grid cell simulations}
\label{subsection: synthetic grid cell simulations}

We simulate populations of noisy grid cells following the same approach used by Redman et al. (2025) \cite{redman2025robust}. We provide a summary below.

An environment of size $L_x \times L_y$ defines the space over which the grid cell patterns exist. In the case of 1D experiments, $L_y = 0$ m. For $N \in \mathbb{N}$ grid cells, the grid spacing and orientation of each grid cell is sampled via $\lambda_i \sim \mathcal{N}(\lambda, \sigma_\lambda^2)$ and $\theta_i \sim \mathcal{N}(\theta, \sigma_\theta^2)$, where $\mathcal{N}(\mu, \sigma^2)$ is a normal distribution with mean $\mu$ and variance $\sigma^2$. Grid phase is sampled as $\mathbf{\phi}_i \sim \mathcal{U}([0, L_x] \times [0, L_y])$, where $\mathbf{\phi}$ is a two dimensional vector, with first component uniformly sampled from $[0, L_x]$ and second component uniformly sampled from $[0, L_y]$. To construct a population with no variability in grid properties, we set $\sigma_\lambda = 0$ m and $\sigma_\theta = 0^\circ$. As noted in the main text, when simulating a 1D environment, $\phi_i = 0$ m and $\theta_i = 0^\circ$, for all $N$ grid cells. 

For each grid cell, we generate an idealized grid response by summing three two-dimensional sinusoids \citep{solstad2006grid}, such that the activity at $\textbf{x} = (x, y) \in [-L_x / 2, L_x / 2]$ $\times  [-L_y / 2, L_y / 2]$ is given by

\begin{equation}
\label{eq: idealized ratemaps}
    G^{(i)}_{\lambda_i, \hspace{0.25mm} \theta_i, \hspace{0.25mm} \phi_i}(\textbf{x}) = G^{\text{max}} \frac{2}{3} \left(\frac{1}{3}\sum_{j = 1}^3 \cos\Big[\textbf{k}_j(\theta_i) (\textbf{x} + \mathbf{\phi}_i) \Big] + \frac{1}{2} \right), 
\end{equation}
where $G^\text{max}$ is the maximal firing rate and $\textbf{k}_j(\theta_i)$ are the wave vectors with $0^\circ$, $60^\circ$, and $120^\circ$ angular differences 
\begin{equation}
\begin{split}
        k_1(\theta_i) &= \frac{k}{\sqrt{2}} \cdot [\cos(\theta_i + \pi / 12) + \sin(\theta_i + \pi / 12), \cos(\theta_i + \pi / 12) - \sin(\theta_i + \pi / 12)] \\
        k_2(\theta_i) &= \frac{k}{\sqrt{2}} \cdot [\cos(\theta_i + 5\pi / 12) + \sin(\theta_i + 5\pi / 12), \cos(\theta_i + 5\pi / 12) - \sin(\theta_i + 5\pi / 12)] \\
        k_3(\theta_i) &= \frac{k}{\sqrt{2}} \cdot [\cos(\theta_i + 3\pi / 4) + \sin(\theta_i + 3\pi / 4), \cos(\theta_i + 3\pi / 4) - \sin(\theta_i + 3\pi / 4)],   
\end{split}
\end{equation}
where $k = 4\pi / (\sqrt{3} \lambda_i)$.

We generated noisy spike rates of $N$ grid cells by assuming a Poisson process and sampling using the idealized ratemaps (Eq. \ref{eq: idealized ratemaps}). More concretely, the activity of grid cell $i$ at position $\textbf{x} = (x, y)$ was assumed to be a random variable with a Poisson distribution, whose mean was $G^{(i)}_{\lambda_i, \hspace{0.25mm} \theta_i, \hspace{0.25mm} \phi_i}(\textbf{x})$. Thus, the probability of observing we sampled $\tilde{G}^{(i)}_{\lambda_i, \hspace{0.25mm} \theta_i, \hspace{0.25mm} \phi_i}(\textbf{x})$ via
\begin{equation}
    \tilde{G}^{(i)}_{\lambda_i, \hspace{0.25mm} \theta_i, \hspace{0.25mm} \phi_i}(\textbf{x}) \sim \mathcal{P}\left[G^{(i)}_{\lambda_i, \hspace{0.25mm} \theta_i, \hspace{0.25mm} \phi_i}(\textbf{x})\right].
\end{equation}

For all numerical experiments, we set $X^\text{max} = 15$. 

\subsection{Deriving the sweet spot}
\label{subsection: Deriving the sweet spot}
The existence of the sweet spot (Sec. \ref{subsection: Sweet spot}) can be found from the theory developed in the main text and in Sec. \ref{subsection: Exact derivation of population vector correlation}) by keeping only the fundamental harmonic (i.e., $k=1$):
\begin{equation}
C(x)\;\approx\;e^{-a x^2} \cos(2\pi x/\lambda),
\end{equation}
where $a= \sigma_1^2 / 4$ and $\sigma_1 = 2\pi \sigma_\lambda / \lambda^2$. 
The peak envelope is $f(x)=e^{-a x^2}$. Its inflection point, where the 
difference of correlation between neighboring peaks is maximal, is the sweet spot $x^*$,
\begin{equation}
f''(x)=2a e^{-ax^2}\big(2ax^2-1\big)=0 
\;\Longrightarrow\; x^* = \frac{1}{\sqrt{2a}} = \frac{\lambda^2}{\sqrt{2}\,\pi\,\sigma_\lambda}.
\end{equation}

We check that this form is consistent with the numerical simulations. For instance, if $\lambda = 0.5$ m and $\sigma_\lambda = 0.02$ m, then $x^* \approx 2.8$ m, which is aligned with what we found in Fig. \ref{fig:1D_sweet_spot}B. Similarly, if we set $\lambda = 0.8$ m and $\sigma_\lambda = 0.05$, then $x^* \approx 2.9$ m, which is aligned with Fig. \ref{fig:distance_decoding}C.

Finally, we note that the maximum slope magnitude is $|f'(x^*)|=\sqrt{2a}\,e^{-1/2}$, so the 
peak-to-peak DoC is
\begin{equation}
\text{DoC}_{\max}\;\approx\;\lambda\,|f'(r^*)|
=\frac{\sqrt{2}\,\pi\,\sigma_\lambda}{\sqrt{e}\,\lambda}
\;\propto\;\frac{\sigma_\lambda}{\lambda}\;, \label{eq:DoC}
\end{equation}
which is consistent with the bands found in Fig. \ref{fig:1D_distance_coding}.
The form of Eq. \ref{eq:DoC} can be understood by \\ 
\begin{align}
    \text{DoC}(n \lambda) &= \frac{1}{2} \Big( \big|C[n \lambda] - C[(n - 1) \lambda]\big| + \big|C[n \lambda] - C[(n + 1) \lambda]\big| \Big) \\
    &= \frac{\lambda}{2} \Big( \big|\frac{C[n \lambda] - C[(n - 1) \lambda]}{\lambda}\big| + \big|\frac{C[n \lambda] - C[(n + 1) \lambda]}{\lambda}\big| \Big). \\
    &\approx \lambda \big| C'[n\lambda] \big| \\
    &\approx \lambda \big| f'[n\lambda] \big|. 
\end{align}

\subsection{Decoding analysis}
\label{subsection: decoding analysis}

Inspired by the rodent behavioral experiment results from Tennant et al. (2018), showing a ``sweet spot'' in distance coding, we performed a decoding analysis on populations of simulated noisy grid cells (Sec. \ref{subsection: synthetic grid cell simulations}). In particular, we sought to explore the kinds of error patterns that emerge when decoding distance from grid cells with heterogeneous grid properties, and to compare these error patterns to those found in the behavioral data. 

We consider a 1D environment of length $10$ m (i.e., $L_x = 10$ m) and a population of $N$ grid cells, whose grid spacing is sampled as $\lambda_i \sim \mathcal{N}(\lambda, \sigma_\lambda^2)$. We then generated $M$ different realizations of the Poisson process, $\tilde{G}^{(i)}_{\lambda_i}(x)$. This gives $\tilde{G}^{(i, j)}_{\lambda_i}(x)$, where $j = 1,...,M$. That is, we have $M$ noisy population vectors, $\tilde{\textbf{G}}^{(j)}_{\lambda}(x)$, for each location in space along the 1D environment. For each of the $M$ population vectors, we compute the correlation function, $C^{(j)}(x)$, which measures the correlation between $\tilde{\textbf{G}}^{(j)}_{\lambda}(x)$ and $\tilde{\textbf{G}}^{(j)}_{\lambda}(0)$, where $x = n\lambda$, $n \in \mathbb{N}$. 

We perform cross-validated distance decoding using the correlation functions. That is, we average $M - 1$ of the correlation functions and then use that as a template to decode distance based on the remaining correlation function. More precisely, we compute the average correlation function, $\widehat{C}(x) = 1/(M - 1) \sum_{m = 1}^{M - 1} C^{(m)}(x)$. Then, for a given integer multiple of the grid spacing $n\lambda$, we assign its decoded location to be \begin{equation}
    \tilde{x} = \min \left\{x \hspace{1mm} \Big| \hspace{1mm} \big| \frac{C^{(M)}(n\lambda)}{\widehat{C}(x)} - 1 \big| < \tau \right\}.
\end{equation}
Assigning the minimum distance where the observed value of $C^{(M)}(n\lambda)$ is within $\tau$ percent of the average correlation, $\widehat{C}(x)$, is inspired by the fact that the rodents in the behavioral experiments tended to stop before the reward zone. This suggests that they stop once evidence has reached a certain threshold. 

In the above description, we have -- for simplicity -- used the first $M - 1$ realizations of the population vector as the ``train'' set and the last correlation vector as the ``test'' set. In practice, we repeated this process $M$ times, each realization being used once as the test set. We take the average error across all $M$ decoded values. We repeat this whole process $K$ times, to sample different populations of grid cells. 

For our numerical experiments, we set $M = 10$, $\tau = 0.05$, $K = 7$, and $N = 64$. We find that the decoding model is sensitive to the choice of $\tau$, although $\tau = 0.025, 0.05, 0.075$ all produce results that are generally consistent with the observed data (Fig. \ref{fig:distance_decoding_vs_threshold}). Future work can explore the plausibility of this decoding scheme in more detail.

\subsection{Analytic 2D extension of theory}
\label{subsection: Analytic 2D extension of theory}
We extend the grid cell activity model to a 2D environment where the grid fields are centered on a regular triangular lattice. Let $\lambda$ represent the grid spacing and $\omega$ denote the width of each isotropic Gaussian field. The two primitive lattice vectors spanning the triangular lattice can be defined as:
\begin{equation}
    \vec{a}_1 = \lambda \begin{pmatrix} 1 \\ 0 \end{pmatrix}, \quad \vec{a}_2 = \lambda \begin{pmatrix} \frac{1}{2} \\ \frac{\sqrt{3}}{2} \end{pmatrix}. \label{eq:primitive_lat_vec}
\end{equation}
Any lattice node $\vec{R}_{\vec{n}}$ is parameterized by the integer pair $\vec{n} = (n_1, n_2) \in \mathbb{Z}^2$ such that $\vec{R}_{\vec{n}} = n_1 \vec{a}_1 + n_2 \vec{a}_2$. Setting the grid orientation $\theta = 0^\circ$ and spatial phase $\vec{\phi} = \vec{0}$ m, the total 2D grid cell activity is expressed as:
\begin{equation}
\label{eq: 2d grid cell equation}
    G_{\lambda, \hspace{0.25mm} \omega}(\vec{x}) = \sum_{\vec{n} \in \mathbb{Z}^2} \exp\left(\frac{-\|\vec{x} - \vec{R}_{\vec{n}}\|^2}{2\omega^2}\right).
\end{equation}
%
Eq. \ref{eq: 2d grid cell equation} is equivalent to a 2D convolution of a Dirac comb with a symmetric Gaussian profile, 
\begin{equation}
    \label{eq: 2d grid cell equation dirac comb}
    G_{\lambda, \hspace{0.25mm} \omega}(\vec{x}) = \operatorname{comb}_{\lambda} (\vec{x}) * g_\omega (\vec{x}),
\end{equation}
where $g_\omega(\vec{x}) = \exp(-\|\vec{x}\|^2 / 2\omega^2)$, $*$ is the spatial convolution operator, and $\operatorname{comb}_{\lambda}(\vec{x}) = \sum_{\vec{n} \in \mathbb{Z}^2} \delta^{(2)}(\vec{x} - \vec{R}_{\vec{n}})$. Now, the Fourier transform becomes,
\begin{equation}
    \label{eq: fourier 2d grid cell activity}
    \widehat{G}_{\lambda, \hspace{0.25mm} \omega}(\vec{\xi}) = \widehat{\operatorname{comb}}_\lambda(\vec{\xi}) \cdot \widehat{g}_\omega(\vec{\xi}),
\end{equation}
where $\vec{\xi} = (\xi_x, \xi_y)$ denotes the spatial frequency vector.

The Fourier transform of a spatial Dirac comb yields another Dirac comb localized on the reciprocal lattice. The primitive reciprocal lattice vectors $\vec{b}_1$ and $\vec{b}_2$ satisfy $\vec{a}_i \cdot \vec{b}_j = 2\pi \delta_{ij}$ and \( |\vec{b}| = \frac{4\pi}{\sqrt{3}\lambda}\), generating the dual triangular frequency lattice nodes $\vec{\mu}_{\vec{k}} = k_1 \vec{b}_1 + k_2 \vec{b}_2$ for $\vec{k} \in \mathbb{Z}^2$. Consequently, Eq. \ref{eq: fourier 2d grid cell activity} can be further realized as
\begin{equation}
\label{eq: expanded fourier 2d grid cell activity}
   \widehat{G}_{\lambda, \hspace{0.25mm} \omega}(\vec{\xi}) \propto  \sum_{\vec{k} \in\mathbb{Z}^2} \delta^{(2)} \left(\vec{\xi} - \vec{\mu}_{\vec{k}}\right) \cdot e^{-\frac{1}{2} \omega^2\|\vec{\xi}\|^2},
\end{equation}
where $\delta^{(2)}(\cdot)$ is the 2D Dirac delta operator. The power spectrum of $\widehat{G}_{\lambda, \hspace{0.25mm} \omega}(\vec{\xi})$ is therefore given by 
\begin{equation}
    \label{eq: power spectrum single 2d grid cell}
    \left| \widehat{G}_{\lambda, \hspace{0.25mm} \omega}(\vec{\xi})\right|^2 \propto
    \sum_{\vec{k} \in\mathbb Z^2} \delta^{(2)} \left(\vec{\xi} - \vec{\mu}_{\vec{k}}\right) \cdot e^{-\omega^2\|\vec{\xi}\|^2}.
\end{equation}

By the Wiener--Khinchin theorem, the spatial autocorrelation function, $R_{\lambda, \hspace{0.25mm} \omega}(\vec{r})$, is the inverse Fourier transform of the power spectrum \cite{Ricker2003-ye}. Taking the inverse Fourier transform of the reciprocal lattice Dirac comb returns the triangular Dirac comb, $\operatorname{comb}_{\lambda}(\vec{r})$. Concurrently, the inverse Fourier transform of the frequency-domain Gaussian envelope $\exp(-\omega^2 \|\vec{\xi}\|^2)$ maps to a spatial Gaussian with a dilated variance. A convenient representation in real space is
\begin{equation}
    \label{eq: 2d autocorrelation}
    R_{\lambda, \hspace{0.25mm} \omega}(\vec{r}) \propto \operatorname{comb}_{\lambda}(\vec{r}) * e^{-\|\vec{r}\|^2/(4\omega^2)} = \sum_{\vec{n} \in\mathbb Z^2} e^{-\frac{\|\vec{r} - \vec{R}_{\vec{n}}\|^2}{4\omega^2}}. 
\end{equation}
That is, the 2D autocorrelation is a sum of isotropic Gaussians centered at each triangular lattice node $\vec{R}_{\vec{n}}$, with each field possessing a spatial width of $\sqrt{2} \omega$. This directly reflects the fact that the autocorrelation of an isotropic 2D Gaussian of width $\omega$ is another 2D Gaussian of width $\sqrt{2}\omega$.

We now consider a population of $N$ grid cells, where each cell $i$ possesses its own distinct grid spacing, $\lambda_i$, and grid orientation, $\theta_i$. We denote the joint collection of these architectural parameters by $\mathcal{P} = \{(\lambda_i, \theta_i)\}_{i = 1}^N$ and maintain the assumption of a fixed field width, $\omega$, across the population. The average population activity is defined as
\begin{equation}
    \label{eq: 2d population grid cell activity}
    G_{\mathcal{P}, \hspace{0.25mm} \omega}(\vec{x}) = \frac{1}{N}\sum_{i = 1}^N G_{\lambda_i, \hspace{0.25mm} \theta_i, \hspace{0.25mm} \omega}(\vec{x}),
\end{equation}
where the individual cell activity $G_{\lambda_i, \hspace{0.25mm} \theta_i, \hspace{0.25mm} \omega}(\vec{x})$ incorporates a standard 2D rotation matrix $R_{\theta_i}$ acting on the underlying triangular lattice vectors, such that the fields are centered at $\vec{R}_{\vec{n}}(\lambda_i, \theta_i) = R_{\theta_i} \vec{R}_{\vec{n}}(\lambda_i, 0)$. This can be achieved by simply rotating the primitive lattice vectors defined in Eq. \ref{eq:primitive_lat_vec}.

We assume that the structural parameters for each cell are independently sampled from a joint distribution $\mathcal{D} = \mathcal{D}_\lambda \times \mathcal{D}_\theta$. Specifically, both the grid spacing and orientation exhibit variability governed by normal distributions:
\begin{equation}
    \lambda_i \sim \mathcal{N}(\mu_\lambda, \sigma_\lambda^2), \quad \theta_i \sim \mathcal{N}(\mu_\theta, \sigma_\theta^2).
\end{equation}
Under these variations, we evaluate the ensemble-averaged (or expected) population value, $\mathbb{E}_\mathcal{D} [G_{\mathcal{P}, \hspace{0.25mm} \omega}(\vec{x})]$. Because the Fourier transform is a linear operation, the expectation operator commutes directly with the transformation, allowing us to evaluate the expected population activity in the frequency domain as $\mathbb{E}_\mathcal{D} [\widehat{G}_{\mathcal{P}, \hspace{0.25mm} \omega}(\vec{\xi})]$. 

Applying this to the individual Fourier components, the rotation in real space induces an identical rotation of the reciprocal lattice in frequency space. The Fourier transform of the population activity thus reads
\begin{equation}
    \label{eq: 2d population fourier transform}
    \widehat{G}_{\mathcal{P}, \hspace{0.25mm} \omega}(\vec{\xi}) = \frac{1}{N}\sum_{i = 1}^N \widehat{G}_{\lambda_i, \hspace{0.25mm} \theta_i, \hspace{0.25mm} \omega}(\vec{\xi}),
\end{equation}
where each constituent term resolves to
\begin{equation}
\label{eq: expanded 2d population fourier component}
   \widehat{G}_{\lambda_i, \hspace{0.25mm} \theta_i, \hspace{0.25mm} \omega}(\vec{\xi}) \propto \sum_{\vec{k} \in\mathbb Z^2} \delta^{(2)} \left(\vec{\xi} - R_{\theta_i}\vec{\mu}_{\vec{k}}(\lambda_i)\right) \cdot e^{-\frac{1}{2} \omega^2\|\vec{\xi}\|^2}.
\end{equation}
Here, $\vec{\mu}_{\vec{k}}(\lambda_i)$ represents the unrotated reciprocal lattice vector scaled by spacing $\lambda_i$, and $R_{\theta_i}$ is the 2D rotation matrix:
\begin{equation}
    R_{\theta_i} = \begin{pmatrix} \cos\theta_i & -\sin\theta_i \\ \sin\theta_i & \cos\theta_i \end{pmatrix}.
\end{equation}
We will, in the next subsection, take $ \vec{\mu}_{\vec{k}}(\lambda_i,\theta_i) = R_{\theta_i}\vec{\mu}_{\vec{k}}(\lambda_i)$. 

\subsubsection{Impact of grid property variability on the 2D Fourier spectrum}

We now consider a population of grid cells with variability in both grid spacing and orientation. We assume the grid parameters are normally distributed. That is, $\lambda_i \sim \mathcal{N}(\lambda, \sigma^2_\lambda)$ and $\theta_i \sim \mathcal{N}(0, \sigma^2_\theta)$. 

Under these variations, the harmonic frequencies (reciprocal lattice vectors) of the Fourier transform, $\vec{\mu}_{\vec{k}}(\lambda_i, \theta_i)$, undergo geometric shifts. For small variabilities ($\sigma_\lambda \ll \lambda$ and $\sigma_\theta \ll 1$ rad), a first-order multivariable Taylor series expansion around the mean spacing $\lambda$ and mean orientation $0$ gives
\begin{equation}
\label{eq: 2d taylor series}
\vec{\mu}_{\vec{k}}(\lambda_i, \theta_i) \approx \vec{\mu}_{\vec{k}}(\lambda, 0) + \frac{\partial \vec{\mu}_{\vec{k}}}{\partial \lambda} \bigg|_{\lambda, 0} (\lambda_i - \lambda) + \frac{\partial \vec{\mu}_{\vec{k}}}{\partial \theta} \bigg|_{\lambda, 0} \theta_i.
\end{equation}
Evaluating the partial derivatives, we find that spacing variability shifts the vector radially by $-\frac{1}{\lambda} \vec{\mu}_{\vec{k}}$, while orientation variability shifts the vector orthogonally by $\vec{\mu}_{\vec{k}}^\perp$ (where $\vec{v}^\perp$ denotes a $90^\circ$ rotation of $\vec{v}$). Hence, to leading order,
\begin{equation}
\vec{\mu}_{\vec{k}}(\lambda_i, \theta_i) \sim \mathcal{N} \left(\vec{\mu}_{\vec{k}}, \Sigma_{\vec{k}}\right),
\end{equation}
where the mean vector is $\vec{\mu}_{\vec{k}} = \vec{\mu}_{\vec{k}}(\lambda, 0)$. Because the radial and tangential shifts are strictly orthogonal, the covariance matrix $\Sigma_{\vec{k}}$ has principal axes aligned with $\vec{\mu}_{\vec{k}}$ and $\vec{\mu}_{\vec{k}}^\perp$, with variances,
\begin{align}
    \sigma_{r, \vec{k}}^2 &= \|\vec{\mu}_{\vec{k}}\|^2 \frac{\sigma_\lambda^2}{\lambda^2} \quad \text{(Radial broadening due to spacing)}, \\
    \sigma_{t, \vec{k}}^2 &= \|\vec{\mu}_{\vec{k}}\|^2 \sigma_\theta^2 \quad \text{(Tangential broadening due to orientation) respectively}.
\end{align}
Thus, each discrete spectral point at $\vec{\mu}_{\vec{m}}$ becomes broadened into a 2D Gaussian defined by the covariance $\Sigma_{\vec{k}}$. Averaging over the continuous distributions of $\lambda_i$ and $\theta_i$ replaces the Dirac delta comb with a sum of normal densities. Defining the squared Mahalanobis distance as $D_{\vec{k}}^2(\vec{\xi}) = (\vec{\xi} - \vec{\mu}_{\vec{k}})^T \Sigma_{\vec{k}}^{-1} (\vec{\xi} - \vec{\mu}_{\vec{k}})$, the expected population Fourier transform yields
\begin{equation}
    \label{eq: expected 2d fourier transform}
    \mathbb{E}_{\mathcal{D}} \left[\widehat{G}_{\mathcal{P}, \hspace{0.25mm} \omega} (\vec{\xi}) \right] \propto \sum_{\vec{k} \in \mathbb Z^2} \exp\left(-\frac{1}{2} D_{\vec{k}}^2(\vec{\xi})\right) \cdot e^{-\frac{1}{2} \omega^2\|\vec{\xi}\|^2}.
\end{equation}
Since we started with small variability, the covariance ellipses are small and non-overlapping. Applying the approximation that for sufficiently narrow spectral peaks, the global field envelope $e^{-\omega^2\|\vec{\xi}\|^2}$ is effectively constant across the width of each peak at $\vec{\mu}_{\vec{k}}$, the expected power distribution becomes,
\begin{align}
    \left|\mathbb{E}_{\mathcal{D}} \left[\widehat{G}_{\mathcal{P}, \hspace{0.25mm} \omega} (\vec{\xi}) \right] \right|^2 &\propto \sum_{\vec{k} \in \mathbb Z^2} \exp\left(-\frac{1}{2} D_{\vec{k}}^2(\vec{\xi})\right)^2 \cdot e^{-\omega^2\|\vec{\xi}\|^2} \\
    &\approx \sum_{\vec{k} \in \mathbb Z^2} \exp\left(-D_{\vec{k}}^2(\vec{\xi})\right) \cdot e^{-\omega^2\|\vec{\xi}\|^2}\\
    &\approx\sum_{\vec{k} \in \mathbb Z^2} \exp\left(-D_{\vec{k}}^2(\vec{\xi})\right) \cdot e^{-\omega^2\|\vec{\mu}_{\vec k}\|^2}
    \label{eq:2dps}
\end{align}


Let $W_{\vec{k}} = e^{-\omega^2\|\vec{\mu}_{\vec{k}}\|^2}$ denote the constant scalar weight for the $\vec{k}$-th peak, and recall that $D_{\vec{k}}^2(\vec{\xi}) = (\vec{\xi} - \vec{\mu}_{\vec{k}})^T \Sigma_{\vec{k}}^{-1} (\vec{\xi} - \vec{\mu}_{\vec{k}})$. 
To find the expected real-space spatial autocorrelation $C(\vec r)$, we inverse Fourier transform the spectrum, term by term:
\begin{equation}
    P_{\vec{k}}(\vec{\xi}) = W_{\vec{k}} \exp\left(-(\vec{\xi} - \vec{\mu}_{\vec{k}})^T \Sigma_{\vec{k}}^{-1} (\vec{\xi} - \vec{\mu}_{\vec{k}})\right).
\end{equation}
Applying the frequency shift theorem and the standard $N$-dimensional Gaussian Fourier transform property, $\mathcal{F}^{-1}\{e^{-\frac{1}{2}\vec x^T A \vec x}\} \propto e^{-\frac{1}{2}\vec r^T A^{-1} \vec r}$ with $A = 2\Sigma_{\vec k}^{-1}$, the inverse transform for a single peak evaluates to:
\begin{equation}
    \mathcal{F}^{-1}\{P_{\vec k}(\vec\xi)\} \propto W_{\vec k} e^{i \vec\mu_{\vec k} \cdot \vec r} \exp\left(-\frac{1}{4} \vec r^T \Sigma_{\vec k} \vec r\right).
\end{equation}
Summing over all reciprocal lattice vectors $\vec{k}$ and grouping the symmetric $\vec{k}$ and $-\vec{k}$ terms (since $\vec\mu_{-\vec k} = -\vec\mu_{\vec k}$ and $\Sigma_{-\vec k} = \Sigma_{\vec k}$), we get:
\begin{equation}
    C(\vec r) \propto \sum_{\vec k \in \mathbb{Z}^2} W_{\vec k} \cos{\left(\vec\mu_{\vec k} \cdot \vec r\right)} \exp\left(-\frac{1}{4} \vec r^T \Sigma_{\vec k} \vec r\right).
\end{equation}
Finally, we evaluate this autocorrelation at the real-space lattice nodes, $\vec{x} = \vec{R}_{\vec{n}}$. The autocorrelation amplitude at the lattice nodes simplifies to:
\begin{equation}
\label{eq:2d_autocorr_nodes}
    C(\vec{R}_{\vec{n}}) \propto \sum_{\vec k \in \mathbb{Z}^2} \exp\left(-\omega^2\|\vec{\mu}_{\vec k}\|^2\right) \exp\left(-\frac{1}{4} \vec{R}_{\vec{n}}^T \Sigma_{\vec k} \vec{R}_{\vec{n}} \right).
\end{equation}
We can again make the following predictions.
\begin{itemize}
  \item \textbf{(P1)} Monotonic decay with radial distance, as $C(\vec{R}_{\vec{n}}) \sim \mathcal{O}\left( e^{-|\vec{R}_{\vec{n}}|^2} \right)$;
  \item \textbf{(P2)} Faster decay with greater variability in both grid spacing and orientation;
  \item \textbf{(P3)} Slower decay for larger $\lambda$, as $C(\vec{R}_{\vec{n}}) \sim \mathcal{O}(e^{-1/\lambda^2})$.
\end{itemize}
Thus, analysis of the correlation function - adapted from the learning paradigms established in \cite{MMM_Dogra_26_1, Thesis_Dogra_2025} - rigorously generates the predictions that underpin our numerical investigations in both the 1D and 2D case. 

\subsection*{Data Availability}

The data analyzed in Sec. \ref{subsection: analysis of 1d grid cell data} was made publicly available by Campbell et al. (2021 \cite{campbell2021distance} and can be accessed at \texttt{https://plus.figshare.com/articles/dataset/VR\_Data\_Neuropixel\_supporting\_Distance-tuned\\\_neurons\_drive\_specialized\_path\_integration\_calculations\_in\_medial\_entorhinal\_cortex\_\\/15041316}. The data analyzed in Sec. \ref{subsection: Sweet spot} was made publicly available by Tennant et al. (2018) \cite{tennant2018stellate} and can be accessed at \texttt{https://datashare.ed.ac.uk/handle/10283/3002}. All simulation and analysis code used in this manuscript is made publicly available at: \texttt{https://github.com/MathePhysics/GriDDE.git}.

\subsection*{Author Contributions}
W.T.R. conceived the study and led the experimental design, neural data analysis, and manuscript preparation. A.S.D. led the development of the mathematical and computational theory. P.D. bridged the computational, mathematical, and neuro-scientific aspects and implemented the numerical experiments. All authors reviewed and contributed to all aspects of the final manuscript.

\subsection*{Acknowledgments}
We thank Michael Goard, Santiago Acosta-Mendoza, and Xue-Xin Wei for helpful discussions on the role of grid property heterogeneity. We thank Kevin Sit for helpful discussion on the implications of the analysis of the experimental results. We also thank Pak Yiu Liu for engaging discussions on the separability of nearby locations. Finally, we thank the anonymous PLOS Computational Biology reviewers whose comments significantly improved the quality of the work.

P. D.'s work is supported by the INSPIRE scholarship from the Department of Science and Technology (DST), under the Ministry of Science and Technology, Government of India. 

A.S.D.’s work is supported by the National Science Foundation under Cooperative Agreement PHY-2019786 (The
NSF AI Institute for Artificial Intelligence and Fundamental Interactions, http://iaifi.org/). A.S.D. was also 
funded by the President’s PhD Scholarships at Imperial College London and by the EPSRC Centre for Doctoral
Training in Mathematics of Random Systems: Analysis, Modelling and Simulation (EP/S023925/1). A.S.D. is also supported by the Center of Mathematical Sciences and Applications (CMSA), Harvard University. 

W.T.R. is supported by the Data Science and AI Institute (DSAI) at Johns Hopkins University. 

\small
\bibliography{main.bib}
\bibliographystyle{unsrt}

\newpage
\beginappendix

\begin{figure}
    \centering
    \includegraphics[width=0.625\linewidth]{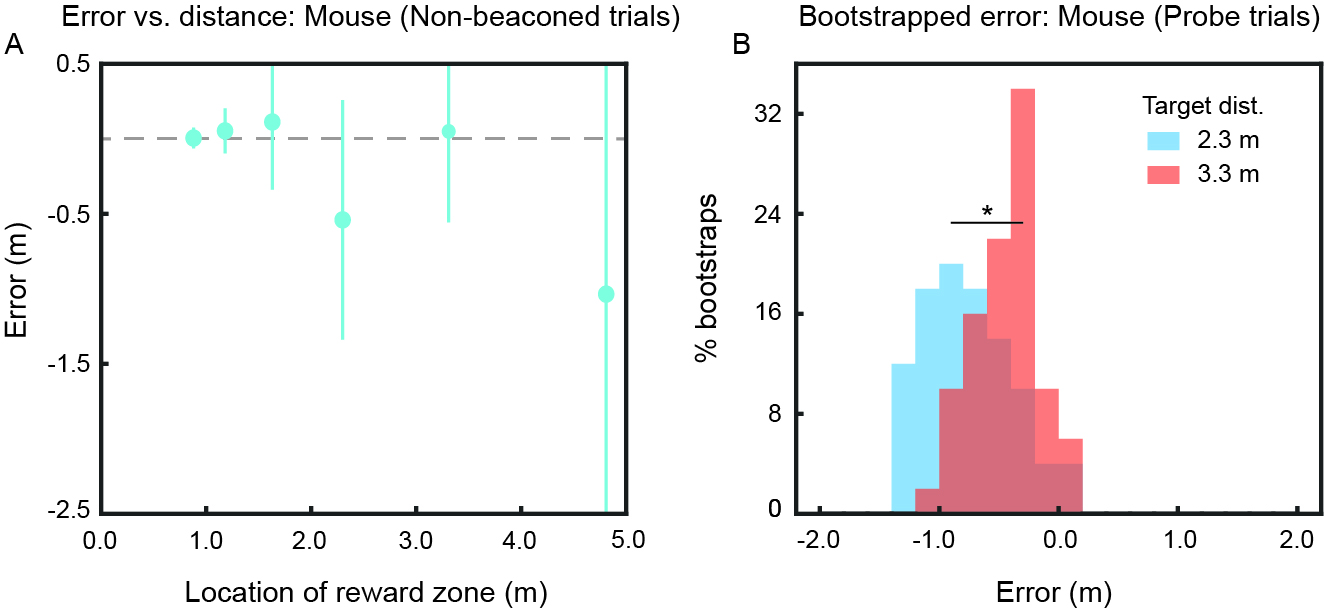}
    \caption{\small \textbf{Additional analysis of Tennant et al. (2018) behavioral experiments \cite{tennant2018stellate}.} (A) Same as Fig. \ref{fig:distance_decoding}B, but for non-beaconed trials. (B) Bootstrapped distribution of average error on trials where the target distance is 2.3 m (blue) and 3.3 m (red). $50$ bootstrapped samples were generated. Star denotes $p-$value $< 0.01$, computed via a two-sample KS test. Medians of the distribution are $-0.82$ m and $-0.40$ m, respectively. }
    \label{fig:statistical_analysis_distance_decoding_data}
\end{figure}

\begin{figure}
    \centering
    \includegraphics[width=0.625\linewidth]{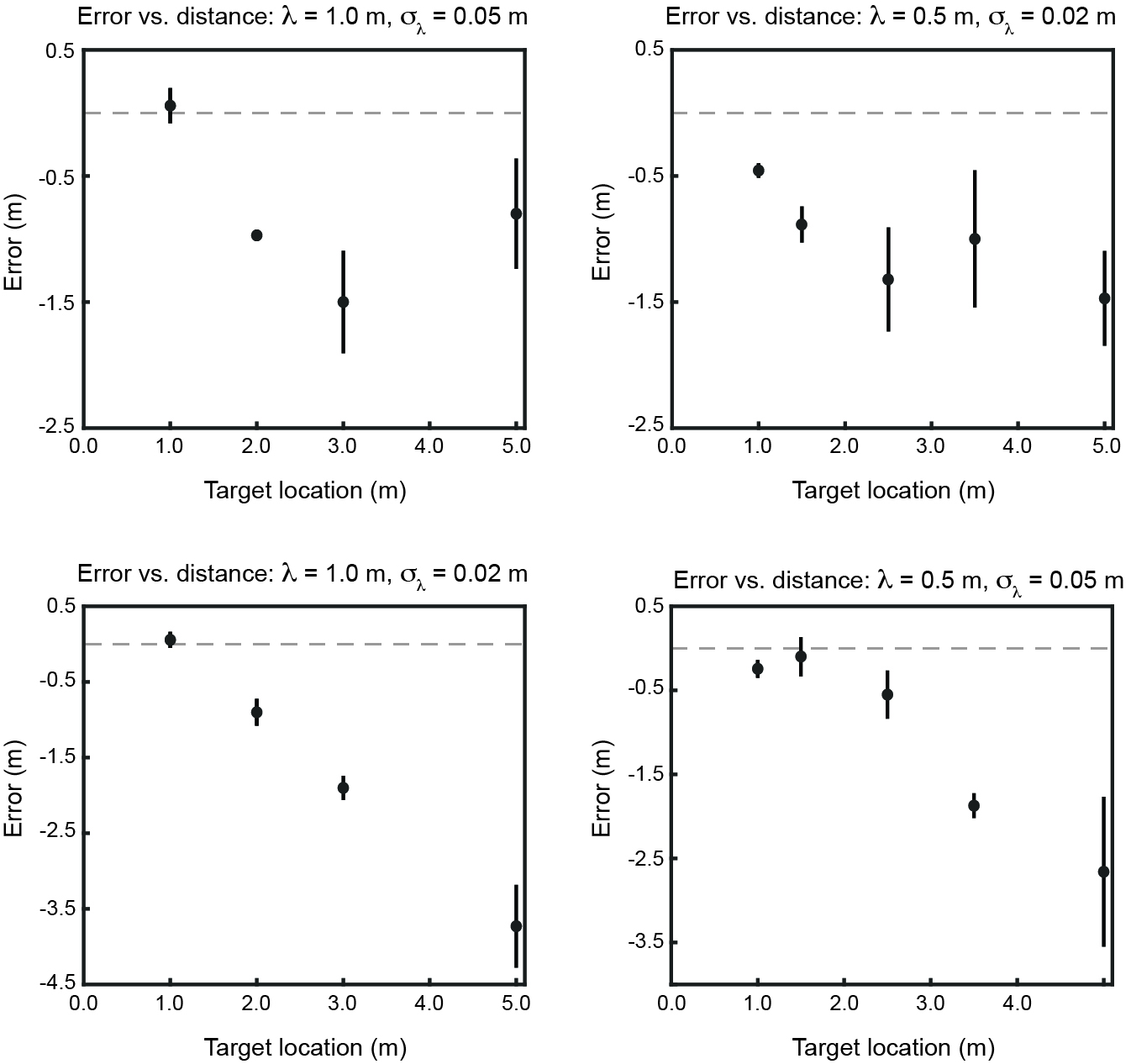}
    \caption{\small \textbf{Decoding error vs. distance in grid simulations for varying values of $\lambda$ and $\sigma_\lambda$.} Same as Fig. \ref{fig:distance_decoding}C, but for more choices of $\lambda$ and $\sigma_\lambda$.}
    \label{fig:distance_decoding_vs_grid_spacing}
\end{figure}

\begin{figure}
    \centering
    \includegraphics[width=0.625\linewidth]{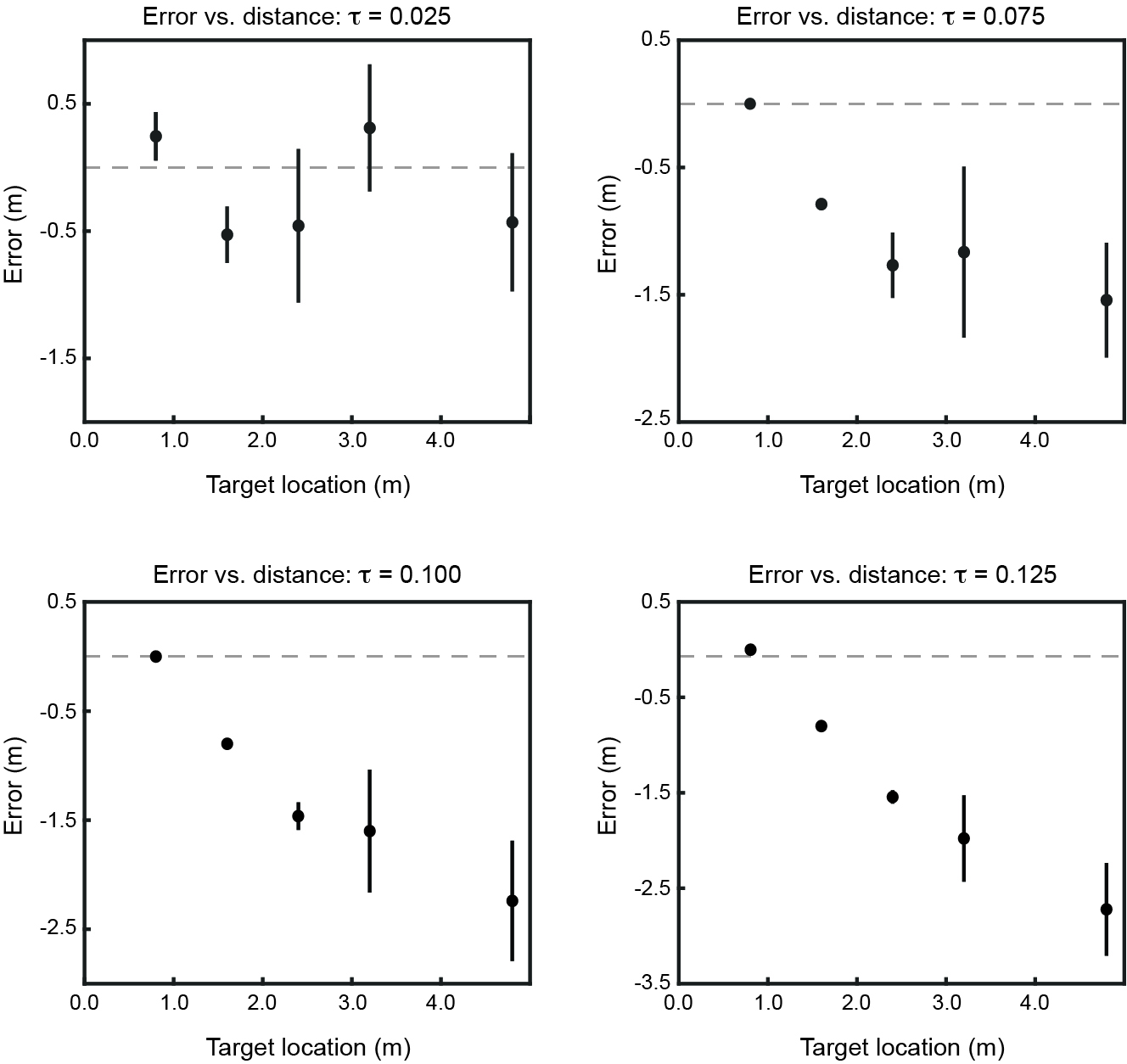}
    \caption{\small \textbf{Decoding error vs. distance in grid simulations for varying values of $\tau$.} Same as Fig. \ref{fig:distance_decoding}C, but for more choices of $\tau$.}
    \label{fig:distance_decoding_vs_threshold}
\end{figure}

\end{document}